\pdfoutput=1
\documentclass[sigconf]{acmart}
\AtBeginDocument{%
  }

\copyrightyear{2025}
\acmYear{2025}
\setcopyright{rightsretained}
\acmConference[CHI EA '25]{Extended Abstracts of the CHI Conference on Human Factors in Computing Systems}{April 26-May 1, 2025}{Yokohama, Japan}
\acmBooktitle{Extended Abstracts of the CHI Conference on Human Factors in Computing Systems (CHI EA '25), April 26-May 1, 2025, Yokohama, Japan}\acmDOI{10.1145/3706599.3720113}
\acmISBN{979-8-4007-1395-8/2025/04}

\usepackage{subcaption}
\usepackage{multirow}
\usepackage{tabularx}
\usepackage{booktabs}
\usepackage[utf8]{inputenc} 
\usepackage[T1]{fontenc}    
\usepackage{hyperref}       
\usepackage{url}            
\usepackage{booktabs}       
\usepackage{makecell}

\usepackage{graphicx} 
\usepackage{caption} 

\begin{document}
\sloppy

\title[LARF]{Let AI Read First: Enhancing Reading Abilities for Individuals with Dyslexia through Artificial Intelligence}

\author{Sihang Zhao}
\affiliation{%
  \institution{The Chinese University of Hong Kong, Shenzhen}
  \city{Shenzhen}
  \country{China}
}
\email{sihangzhao@link.cuhk.edu.cn}

\author{Shoucong Carol Xiong}
\affiliation{%
  \institution{Zhejiang University}
  \city{Hangzhou}
  \country{China}}
\email{carolhsiung@zju.edu.cn}

\author{Bo Pang}
\affiliation{%
  \institution{Chinese Academy of Science}
  \city{Beijing}
  \country{China}}
\email{bopang@cnic.cn}

\author{Xiaoying Tang}
\affiliation{%
  \institution{The Chinese University of Hong Kong, Shenzhen}
  \city{Shenzhen}
  \country{China}}
\email{tangxiaoying@cuhk.edu.cn}


\author{Pinjia He}
\authornotemark[1]
\affiliation{%
  \institution{The Chinese University of Hong Kong, Shenzhen}
  \city{Shenzhen}
  \country{China}}
\email{hepinjia@cuhk.edu.cn}



\renewcommand{\shortauthors}{Zhao et al.}

\begin{abstract}
Dyslexia, a neurological condition affecting approximately 12\% of the global population, presents significant challenges to reading ability and quality of life. Existing assistive technologies are limited by factors such as unsuitability for quiet environments, high costs, and the risk of distorting meaning or failing to provide real-time support. To address these issues, we introduce LARF (Let AI Read First), the first strategy that employs large language models to annotate text and enhance readability while preserving the original content. We evaluated LARF in a large-scale between-subjects experiment, involving 150 participants with dyslexia. The results show that LARF significantly improves reading performance and experience for individuals with dyslexia. Results also prove that LARF is particularly helpful for participants with more severe reading difficulties. Furthermore, this work discusses potential research directions opened up by LARF for the HCI community.

\end{abstract}


\begin{teaserfigure}
  \includegraphics[width=\textwidth]{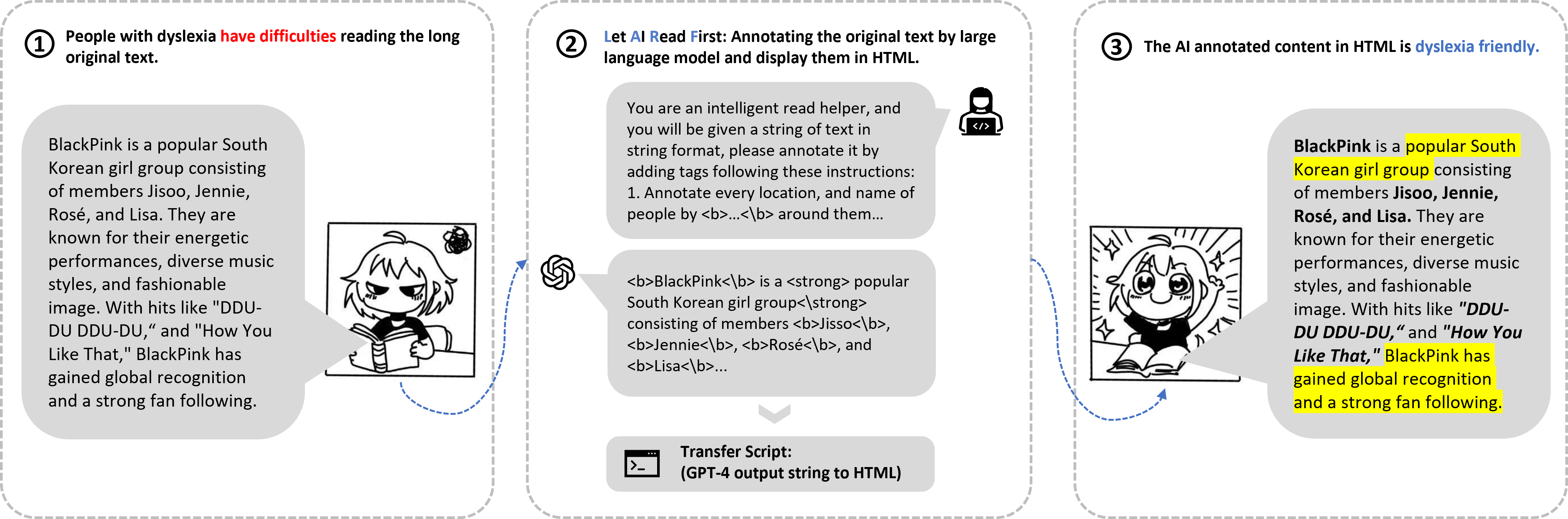}
  \caption{People with dyslexia always have difficulties while reading. We propose a method Let AI Read First (LARF) that uses language models to annotate the original text and display them in HTML format. Our experiment validates that LARF can improve reading performance and improve the reading experience for individuals with dyslexia.}
  \Description{The image is arranged in three labeled sections, illustrating a process of converting a long text into a more accessible format for people with dyslexia: Left Section (Step 1) A heading in red text reads: “People with dyslexia have difficulty reading the long original text. Below it is a text box containing a paragraph about BlackPink, stating: “BlackPink is a popular South Korean girl group consisting of members Jisoo, Jennie, Rosé, and Lisa. They are known for their energetic performances, memorable music, and stylish image. With hits like ‘DDU-DU DDU-DU’, ‘Kill This Love’, and ‘How You Like That’, BlackPink has gained global recognition and a strong fan following.” To the right of the text box, there is a small cartoon figure (in black and white style) standing and speaking or pointing, illustrating the idea of the user or a narrator. Middle Section (Step 2) A heading reads: “Let it Read First: Annotating the original text by large language model and displaying them in HTML.” A smaller text block shows an HTML-like snippet. It uses HTML tags (e.g., <strong> and <p>) around the text about BlackPink. The same cartoon figure appears here, now depicted more prominently and looking as if it is giving instructions or performing an action. Right Section (Step 3) A heading reads: “The AI annotated content in HTML is dyslexia friendly.” Below, there is a revised text in color, which highlights that BlackPink “gained global recognition and a strong fan following.” The rest of the text is broken up into smaller chunks, presumably making it easier to read. The same cartoon character is shown again, indicating the final step in the process. Arrows or lines connect these three sections, emphasizing the workflow: from the original long text (Step 1), to the annotated version in HTML (Step 2), and finally to a dyslexia-friendly output (Step 3).}
  \label{fig:teaser}
\end{teaserfigure}

\received{20 February 2007}
\received[revised]{12 March 2009}
\received[accepted]{5 June 2009}

\maketitle
\section{Introduction}
\renewcommand{\thefootnote}{}
\footnotetext{*Pinjia He is the corresponding author.}
Dyslexia is a neurodevelopmental impairment that affects reading abilities, typically manifested as challenges to reading fluency, speed, and comprehension. Approximately 12\% of the global population has dyslexia \cite{TP-toolbox-web}. Individuals with dyslexia often struggle with word decoding and recognition, which also affects their comprehension, fluency, and vocabulary. Current interventions, mainly in the form of accessible designs, tend to focus on only a few areas: converting text to speech \cite{laga2006kurzweil}, videos or games \cite{larco2021moving, ostiz2021improving}, adjusting text font through electronic readers \cite{rello2012layout, rello2013good} (e.g., character size, colour, spacing between words), and replacing complex words with simpler synonyms \cite{rello2014evaluation}. Nevertheless, these efforts often exhibit one or more of the following limitations: (1) In scenarios demanding quiet, such as conferences and exams, the use of multimedia-assisted tools presents practical difficulties. (2) Converting text descriptions into videos or games manually can be both expensive and non-real-time. (3) Simple synonym substitution and rewriting may alter the original meaning, rhymes or emotions of the original texts. Compared to the available knowledge about reading difficulties and the demonstrated capabilities of AI models, there are relatively few accessible designs that effectively address these challenges \cite{mccarthy2010we}. With the rapid development of AI \cite{bommasani2021opportunities}, numerous spelling assistance tools for dyslexia have demonstrated considerable capabilities \cite{galuschka2020effectiveness, rello2014computer, goodman2022lampost}. However, we have not yet discovered any existing reading assistance tools or research that has utilised or discussed how to integrate state-of-the-art AI techniques to address these issues in assistive reading tools for people with dyslexia.

Therefore, to fill these gaps, we propose an AI-based presentation strategy to assist people with dyslexia in reading. We introduce LARF (\textbf{L}et \textbf{A}I \textbf{R}ead \textbf{F}irst), the first AI-based method that annotates ``important'' information in texts with highlights, bolding, underlining, and other marks. This approach aims to help readers focus more easily on the key content of the original text, thereby enhancing their reading performance and experience. Unlike direct AI-generated summaries, LARF's design of annotating the original text preserves the maximum amount of original textual information. 

Our main hypothesis is that LARF can improve the overall reading performance and experience of people with reading difficulties. Consequently, we conducted a large-scale experiment (N = 150) to evaluate this hypothesis. Participants self-reported having or likely dyslexia and having English as their mother tongue. They were randomly assigned into three groups: a control group that read the original reading materials directly, a conventional group in which participants read the same materials processed by Bionic Reading \cite{BionicReading}, and a LARF group that the reading materials annotated by OpenAI's GPT-4 \cite{ChatGPT}. Using multiple-choice questions, we tested the accuracy in recalling, retrieving details, and comprehension levels. The experimental results show that participants who read GPT-4 annotated texts demonstrate better reading performance compared to those using traditional methods or in the control groups. Participants were also asked to complete a series of subjective evaluations to assess their user experience with LARF or the conventional tool. The results indicate that GPT-4 annotated texts significantly improve perceived user-friendliness, overall satisfaction, perceived helpfulness, future use, and recommendation tendencies. Users also believed that this method should be applied as a text presentation method for dyslexic populations in more scenarios (e.g., exams, accessible website design). 

\section{Related Work and Background}
In the realm of accessible design interventions to alleviate reading difficulties, myriad solutions have been proposed. A popular trend has been to incorporate text-to-speech conversion \cite{laga2006kurzweil}, enabling individuals with reading difficulties to access written content orally. Parallelly, innovative efforts have been made to employ multimedia elements such as videos and games to facilitate reading comprehension \cite{larco2021moving}. However, these software solutions are often limited by contextual restrictions, as text-to-speech conversion becomes impractical in settings that require silence, such as conferences or exams.
Moreover, despite proven effectiveness \cite{cullen2013effects, stodden2012use}, the high cost of software like Kurzweil3000 limits its widespread adoption \cite{cullen2013effects}. Furthermore, traditional methods of transforming textual information into images, audio, or even games require substantial involvement from experienced annotators, developers, and designers. This significantly escalates costs and eliminates the possibility of real-time use, thus further restricting its application scenarios. The other trend is using adjustable text presentation, allowing for modifications in character size, colour, and word spacing \cite{rello2012layout, rello2013good}. Santana et al. created Firefixia, which is a browser extension that enables dyslexic readers to tailor websites for enhanced readability \cite{de2013firefixia}. Text4All \cite{topac2012development}, an online service for web pages, and the Android IDEAL eBook reader4 for e-books are customisation tools informed by previous research in dyslexic individuals \cite{rello2012layout}. Text4All extends its offerings to include medical language adaptation, terminology annotation, and language analysis. Currently, a popular method called Bionic Reading \cite{BionicReading} revises texts so that the most concise parts of words are highlighted. This guides the eye over the text, and the brain remembers previously learnt words more quickly. Although these methods can be applied in a broader range of contexts, they treat all text as a uniform entity, lacking a targeted emphasis on key segments such as definitions or summary sentences. This results in substantial room for improvement to improve reading performance and experience. In another approach, complex words are replaced with simpler synonyms to aid comprehension \cite{rello2014evaluation}. However, such an approach not only fails to guarantee accuracy in the context of substitution (i.e., it may completely distort the original intent of the text) but may also affect the literary attributes of the text, such as emotional intensity and rhythm. 

Niklaus et al. evaluated the digital reading rulers and found that digital rulers can help people with dyslexia better focus on the text and improve their reading speed \cite{niklaus2023digital}. Li et al. suggested Reader View websites with low visual complexity can benefit the reading performance and user experience of people with and without dyslexia \cite{li2019impact}. Despite the wealth of knowledge surrounding reading difficulties, traditional accessible designs addressing these challenges remain limited \cite{mccarthy2010we}. Considering the rapid advancements in AI, the incorporation of AI models with superior reading comprehension and creativity into accessible design offers a promising area for further exploration. As these models become increasingly versatile and powerful, their intersection with accessible design presents a promising opportunity to overcome the limitations of current solutions.

\section{Method and Data}
\subsection{Workflow of LARF}


The workflow of LARF is illustrated in Fig. \ref{fig:teaser}, in which LARF takes the original text as input in string format. Guided by preset prompts, GPT-4 processes the original text, incorporating the Hyper Text Markup Language (HTML) tags \cite{berners1990worldwideweb, berners1994world}, which can be used to manipulate the display of text, such as ``bold," ``highlighting,'' ``italics,'' changing font colour, and adjusting font size. Subsequently, a Python script compiles this HTML-tagged string into an HTML file, serving as the final output. Consequently, users receive a presentation where specific information has been modified with bold formatting or highlighting, while the textual content remains entirely unchanged. The simple example (Fig. \ref{fig:teaser}) shows a segment taken from Wikipedia about BlackPink \cite{blackpink2024}. GPT-4 was asked to highlight sentences that serve a summarizing role using \textless mark\textgreater \textless \textbackslash mark\textgreater tags and to bold important names and items using \textless b\textgreater \textless \textbackslash b\textgreater tags. After processing the output of GPT-4 with the transfer scripts, the user gets the GPT-4-annotated content shown on the right-hand side.


In the subsequent experiment, we adjusted the prompts by using different labels, thereby modifying the presentation of the text. The detailed default prompts can be found in Appendix \ref{default prompt}.

\subsection{Data}
In the experiment, we processed the reading materials using GPT-4 API. We also used GPT-4 together with human evaluation to score the participants' short-answer responses in the subsequent experiment. The version of GPT-4 is the ChatGPT July 20 version, with the temperature set to 0 to ensure reproducibility of results. All the specific prompts and generation logs can be found in the supplement material and Appendix \ref{default prompt}. We employed the Bionic Reading \cite{BionicReading} as a representative of conventional tools to process the corpora in subsequent experiments, as it is one of the most widely used reading performance improvement solutions. Existing research suggests that Bionic Reading can improve students reading proficiency \cite{ariyani2023improving}. This tool includes two key parameters: “Fixation,” which determines the expression of letter combinations, set to the default value of 3 (ranging from 1 to 5), and “Saccade,” which controls the visual jumps between fixations, set to the default value of 10 (ranging from 10 to 50). In this paper, we also apply the default value of 10. The example of Bionic reading can be found in Appendix \ref{Bionicreading}.

\section{Ethic \& Transparency}
This experiment was approved by the Institutional Review Board (IRB) of our affiliation. All participants were recruited through Prolific, an online research platform \cite{prolific}. To ensure data protection and confidentiality, participants were informed that their responses would be anonymized, with all identifiable information removed before analysis. Additionally, the survey (including confidentiality information), raw experimental data, GPT-4 processing history (including evaluations and annotations), and data analysis code are available in the supplementary materials. Examples of our prompts and questions in the questionnaires are provided in Appendix \ref{detail}. The LARF demo is publicly available for free trials at \href{https://github.com/LARF2025/LARF-CHI-EA-25}{https://github.com/LARF2025/LARF-CHI-EA-25}.

\begin{figure*}[htbp]
  \centering
  \includegraphics[width=0.95\textwidth]{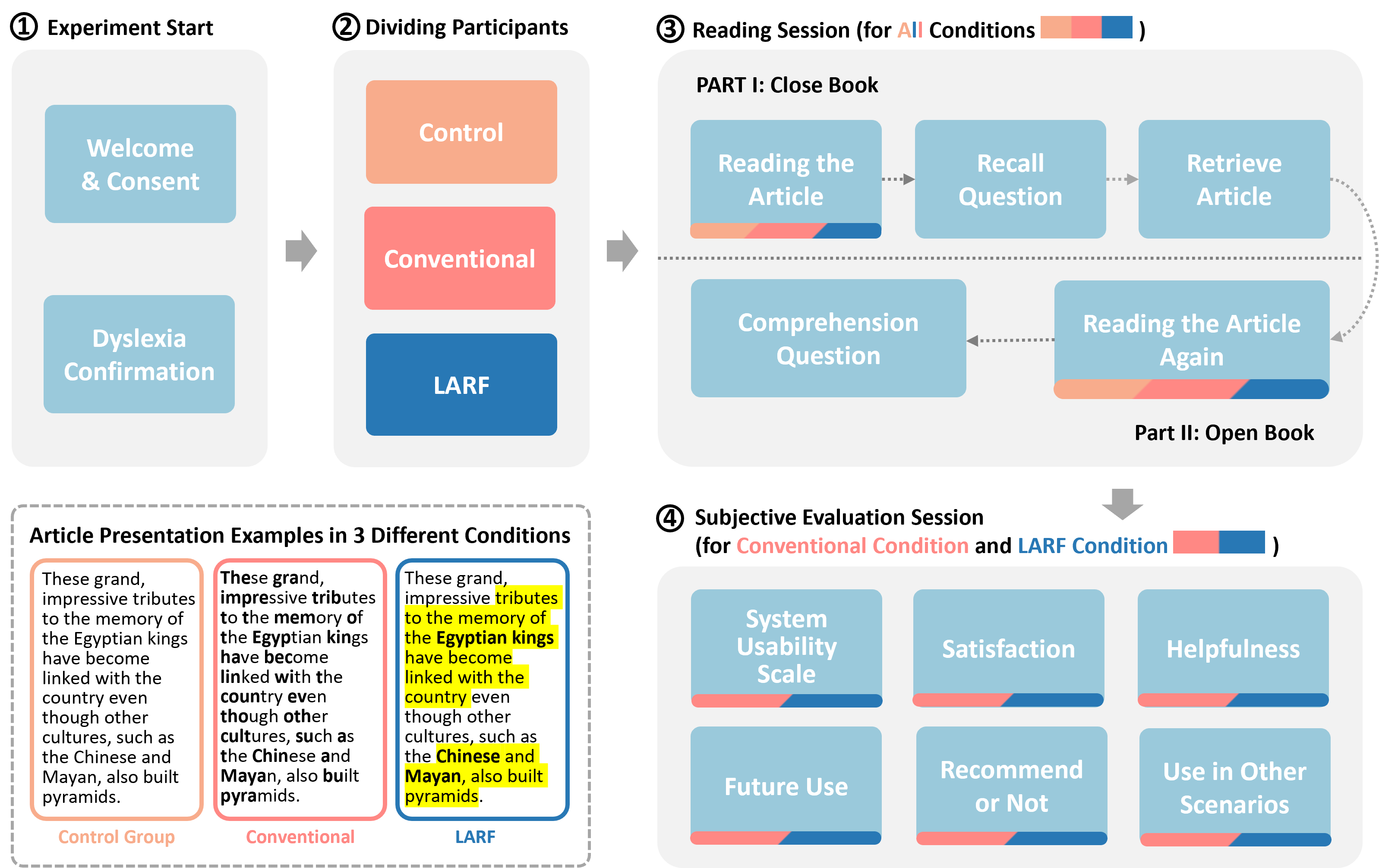}
  \Description{This figure is a four-step flowchart illustrating the experimental procedure and grouping:

1. In the top-left box (Step 1), participants go through "Welcome & Consent" and "Dyslexia Confirmation."  
2. They are then randomly assigned to one of three groups—Control (pink), Conventional (orange), or LARF (blue)—as shown in the middle (Step 2).  
3. Next (Step 3), the flow proceeds to two parts of the experiment: "PART I: Close Book," which involves reading the article, answering recall and comprehension questions, retrieving the article, and then reading it again, followed by "PART II: Open Book." These steps are depicted in multiple connected boxes on the right.  
4. Finally (Step 4), the lower section of the figure shows example reading materials or annotations for each group. The Control Group (on the left) has standard text with no special formatting. The Conventional Group (in the middle) and the LARF Group (on the right) have different highlighting or bold label styles, illustrated by colored blocks of text. The rightmost portion also lists subjective evaluation metrics (e.g., System Usability Scale, Satisfaction, Helpfulness, Future Use, Recommendation, and Other Use Scenarios) for the Conventional and LARF conditions.}
  \caption{Experiment Procedure. Participants are randomly assigned to three conditions, and then they are asked to finish the reading session. They are required to read the same article but with different presentations. Participants in the conventional condition group and LARF condition group are required to finish a subjective evaluation session after they finish the reading session.}
  \label{mainfig}
\end{figure*}

\section{Experiment}
\subsection{Experiment Setup}
Our experiment focused on English language reading. We chose Reading Test 115, Passage 2, “The Step Pyramid of Djoser,” a descriptive and factual reading text from the IELTS \cite{IELTS} Academic as the corpus in this study. This decision was motivated by the comprehensive nature of the IELTS Academic reading test, which employs a long-form format featuring texts sourced from books, journals, magazines, and newspapers \cite{abbott2020cambridge}.  The IELTS Academic test is equipped with expertly formulated questions and standardized answers, which further enhance the reliability and validity of our study. Given these qualities, the IELTS Academic test serves as an ideal tool for assessing adult reading performance.

\subsection{Method and Experiment Procedure}

We recruited 150 participants ($M_{age} = 36.8$; 33.3\% female) from Prolific \cite{prolific}, an online research platform. All participants either had a medical diagnosis of dyslexia, were undergoing a diagnostic process, or strongly suspected they had undiagnosed dyslexia. Additional demographic details are provided in Appendix \ref{participants}. Participants were randomly assigned to one of three experimental conditions: control (unmodified text), conventional tool (Bionic Reading), or LARF (GPT-4 annotations). In the LARF condition, participants were not informed that GPT-4 had produced the annotations, in order to minimize any psychological priming effects.

The study began with participants completing a Dyslexia Checklist (refer to Appendix \ref{Dyslexia Checklist}), designed to assess the severity of various reading-related challenges they face based on personal experiences. Afterwards, they read an article and answered a series of recall questions to evaluate their retention of key details, such as the main character’s name and aspects of a described pyramid. Our design included six recall questions, alongside an attention check question (refer to Appendix \ref{Recall Question}). Following the recall task, participants were asked to retrieve as many details from the article as possible. Then, the same article was presented again, immediately followed by a reading comprehension assessment on the same page. After reading and finishing the reading comprehension assessment, participants in the control condition provided demographic information (age, gender, educational background) and completed the experiment. Participants in the conventional tool and LARF conditions also evaluated the modifications and annotations made by these tools. We used an adapted version of the System Usability Scale \cite{brooke1996sus} to assess tool usability. Participants then rated the tool’s perceived helpfulness, satisfaction, intention to continue using it, and likelihood of recommending it to others. Participants in the conventional tool condition completed the experiment after providing demographic information. In contrast, participants in the LARF condition were asked about their preference for a personalized LARF tool before providing demographic information. The experimental procedure and session details are shown in Fig. \ref{mainfig}.

\section {Result and Analysis}
\label{ex1}
\subsection{Attention Check and Dyslexia Checklist}
Of the initial 150 participants, 2 failed to pass the attention check and were consequently excluded from further analysis. The remaining 148 participants were included in subsequent analyses. There are 51 participants in the control condition, 49 in the conventional tool condition, and 48 in the LARF tool condition. The detail of the attention check is given in the Appendix \ref{attention}. Before reading the article, participants assessed their own dyslexia levels using the Dyslexia Checklist (see Appendix \ref{Dyslexia Checklist} for Dyslexia Checklist). This checklist comprises six items that evaluate comprehension issues, word recognition difficulties, decoding difficulties, memory problems, attentional difficulties, and visual disturbance. We calculated the average scores from these items to determine each participant’s overall dyslexia level (Cronbach’s alpha = 0.91). Statistical analysis reveals no significant differences in dyslexia levels across the three conditions ($M_{control}=3.80$, $SD = 1.65$; $M_{conventional}=3.48$, $SD = 1.41$; $M_{LARF}=3.49$, $SD = 1.29$; $F(2, 145) = .755$, $p = .472$), which indicates that participants are balanced among three conditions

\subsection{Reading Time}
\begin{figure*}[htbp]
  \centering
  \includegraphics[width=1\textwidth]{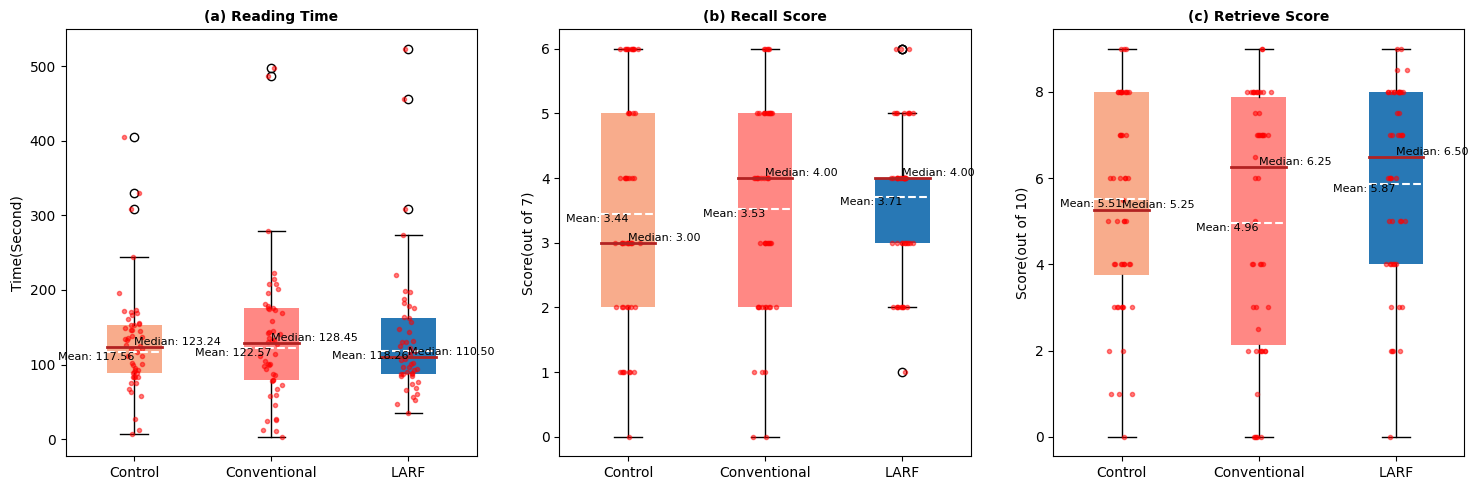}
  \Description{The figure comprises three side-by-side boxplots labeled (a), (b), and (c), each comparing three groups (Control, Conventional, and LARF). Subfigure (a) displays Reading Time (in seconds) on the y-axis, with boxplots and scattered points indicating individual data. The Control group has a wider spread with higher outliers (around 400–500 seconds), while the LARF group shows a lower median (around 70 seconds). Subfigure (b) shows Recall Score on a 0–7 scale, where the Control and Conventional groups have medians near 3, and the LARF group has a higher median (about 4). Subfigure (c) presents Retrieve Score on a 0–10 scale, with the LARF group again scoring higher (median around 6) compared to the Control and Conventional groups (medians around 5). In each boxplot, the box covers the interquartile range (Q1–Q3), the horizontal line inside the box represents the median, whiskers extend to the minimum and maximum data within 1.5 times the interquartile range, and any circles outside are outliers. The overall pattern suggests that participants in the LARF group read faster and achieved higher recall and retrieve scores than those in the other two groups.}

  \caption{(a) shows the differences in reading time under three different conditions. Though the pattern is not significant, we can observe that users in the LARF group do less ``glance over and skip the article.'' Furthermore, their overall reading time is more concentrated in areas with shorter durations. Subfigure (b) and (c) respectively represent the scores of users in the retrieve and recall phases. It can be observed that compared to other groups, participants reading the LARF-marked texts exhibit better recall ability (marginally significant) and a superior capability to remember the details of the articles (significant).}
  \label{fig: Fig. 9}
\end{figure*}

Eight participants are identified as outliers based on their initial reading times, defined as reading times $>$ Q3 + 1.5 $\times$ IQR or $<$ Q1 - 1.5 $\times$ IQR, and were thus excluded from this part of the analysis. Consequently, the final analysis on reading time was conducted with 138 participants. Covariates, including education, age, gender, and dyslexia level, were accounted for in the analysis. We introduced the education level, age, gender, and dyslexia level as covariates in our one-way ANOVA analysis. It shows no significant differences in reading times across conditions ($M_{control} = 117.56, SD = 46.47; M_{conventional} = 122.57, SD = 62.70; M_{LARF} = 118.26, SD = 50.69; F(2, 129) = .160, p = .853$). However, considering the length of the corpus (338 words), and average reading speed of 238 words per minute for English readers \cite{brysbaert2019many}, those who spent less than 30.2 seconds (0.05 quantile) were considered relatively impatient. Fig. \ref{fig: Fig. 9}(a) shows that participants using LARF did not fall below 30 seconds and were concentrated within a shorter, reasonable range. This suggests that LARF may aid in attracting user attention and enhancing reading patience and confidence.

\subsection{Recall and Retrieve Performance}
In the recall section, participants answered six questions, with one point awarded for each correct response (example questions can be found in Appendix \ref{Recall Question}). The analysis included education, age, gender, dyslexia level, and reading times as covariates in a one-way ANOVA. As shown in Fig. \ref{fig: Fig. 9}b, \textbf{participants in the LARF condition tended to score higher (7.8\% higher than the control group and 5.1\% higher than conventional group) than the other two conditions}, though the difference was not statistically significant ($M_{control} = 3.44, SD = 1.74; M_{conventional}=3.53, SD = 1.64; M_{LARF} = 3.71, SD = 1.20; F(2, 128) = .303, p = .739$).

In the retrieve section, participants were instructed to retrieve as many details from the article as possible (example question can be found in Appendix \ref{Retrieve Question}.) We employ GPT-4 to evaluate the quality of participants’ retrieval performance, utilizing a scoring range of 0 to 10. The assessment scores of GPT-4 underwent verification by two human reviewers, each of whom independently cross-checked the scores. The reviewers made only one significant correction to the scores, which was clearly erroneous. The scoring criteria can be found in Appendix \ref{Criteria} and the GPT-4 score logs are available for reference in the supplementary materials. A similar one-way ANOVA analysis was conducted. The results in Fig. \ref{fig: Fig. 9}c clearly show a significant difference across three conditions ($F(2,128) = 3.465, p = .034$). \textbf{Participants in the LARF condition ($M_{LARF} = 5.87$, $SD = 2.30$) scored higher (6.5\% higher than the control group and 18.3\% higher than the conventional group) than the other two conditions }($M_{control} = 5.51$, $SD = 2.38$, $M_{conventional} = 4.96$, $SD = 2.89$).

\subsection{Comprehension Performance}
Comprehension performance was assessed using a similar method in the IELTS examination. Participants were required to identify the correct two statements out of six that were presented in the article. To ensure accuracy in scoring, participants selecting more than two statements were automatically assigned a score of zero, as per our predefined criteria that only two statements were correct. Our analysis revealed that 72.92\% of participants in the LARF condition correctly chose the exact two statements. In contrast, this accuracy was observed in 64.71\% of participants in the control condition and 67.35\% in the conventional condition. Additionally, we evaluated whether participants were able to identify at least one correct statement. In this regard, 100\% of participants in the LARF condition succeeded in choosing at least one correct statement, whereas the corresponding figures were 92.16\% for the control condition and 87.76\% for the conventional condition. We conservatively believe that this indicates \textbf{LARF can to some extent enhance the participants' reading comprehension skills.}


\subsection{Subjective Evaluation}
We conducted a separate analysis to compare the subjective evaluations of the annotation tools between the conventional and LARF conditions. The questionnaire items and corresponding results are presented in Appendix \ref{Subjective Evaluation Result in Experiment 1} Fig. \ref{fig: Fig. 10}. \textbf{Overall, participants in the LARF condition rendered more favourable evaluations than those in the conventional condition.} Notably, participants exposed to LARF-generated annotations reported more positive perceptions and future behaviour tendencies across multiple dimensions. The detailed questions and results are shown in Table \ref{tab:sbj_eva} and Table \ref{tab:sbj_eva_2} in Appendix \ref{sub_1}. \textbf{The result suggests that participants in the LARF group show more overall satisfaction, they also reported that LARF is more helpful and easier to use compared to Bionic Reading.}

\subsection {Post Hoc Evaluation}
 People with dyslexia often experience different subsets of challenges \cite{morris2018understanding}. Given the varying severity of dyslexia among participants, resulting in distinct reading challenges, we conducted a post hoc evaluation to assess LARF's efficacy across different degrees and categories of reading difficulties, focusing on its effects on various symptoms of reading disabilities. Based on previous research, we calculated the mean ± 1 standard deviation (SD) for each dyslexia item. Participants whose self-reported dyslexia scores were higher than M + 1 SD were classified as having severe dyslexia, while those with scores lower than M – 1 SD were classified as having mild dyslexia.\textbf{ Results indicate that LARF is especially helpful for participants with severe dyslexia.} As Fig. \ref{fig:subfig comprehension} in Appendix \ref{Supplemental Figures} shows, the improvement in participants' reading performance is more pronounced in those with severe dyslexia. Similar results can also be observed in their recall (Fig. \ref{fig:subfig recall}) and retrieval (Fig. \ref{fig:subfig retrieve}) performance.

\section{Discussions}
\label{future work}
Based on our theoretical foundation and software demonstration (refer to Appendix \ref{software demo}), as shown in Fig. \ref{larf scenarios}. LARF can be applied to various smart scenarios (e.g., PCs, tablets, and VR), yet high GPU requirements for LLM inference remain a challenge. Exploring smaller models that maintain annotation quality is thus crucial. From an HCI perspective, future work could investigate how best to present AI-generated annotations (e.g., highlight length, colour, or font) for users with reading difficulties, as well as the potential long-term effects on memory and learning. LARF may also be extended to subtitles in videos or live streams, although the impact on neurodiverse populations (such as individuals with ADHD) calls for further exploration. Design solutions should offer customizable annotation settings and integrate seamlessly with existing accessibility features; voice or gesture controls may be essential for VR or compact devices. To ensure privacy, local or end-to-end model inference is preferred, supported by model fine-tuning and well-crafted prompts to enhance annotation quality. Additionally, we have an interesting finding: compared to the control group, Bionic Reading does not appear to improve users' reading performance. In related studies published later than our experiment, they had the similar conclusion \cite{snell2024no}.

\section{Limitation}
During our experiments and software development, we faced several limitations. Considering the experiment cost and participants' patience, We kept comprehension and recall tasks relatively simple and limited the number of questions. However, some questions may have been “too easy,” resulting in some observed patterns without clear statistical significance. We also decided not to include a “random labelling group” to account for placebo effects, though we believe such effects would be minimal. As mentioned in Section \ref{future work}, this study does not investigate the interaction between different annotation types nor determine which is most beneficial. We likewise did not explore how to select optimal default prompts for user engagement. While changing font size and colour in HTML is possible, we have not addressed it here. Previous work \cite{dickinson2002ongoing} indicates that letting users set their own preferences can improve reading accuracy, so in our demo users can specify which information they want GPT-4 to annotate and how it should appear. Nevertheless, in our experiment, participants only used the default prompt. For BionicReading, we set the default parameter, and in real-world usage, users can also customise the settings.

\section{Conclusion}
We introduce LARF, an AI-annotated text approach designed to enhance the reading abilities of individuals with dyslexia. Our Experiment (N=150) validated LARF's effectiveness in improving dyslexic readers' performance and experience, including recall of details, reading comprehension efficiency, and engagement, outperforming the conventional technique.

\begin{acks}
We used LLMs to enhance the linguistic precision and coherence of the content.
This work is funded by Guangdong Basic and Applied Basic Research Foundation (No. 2024A1515010145) and Shenzhen Science and Technology Program (No. ZDSYS20230626091302006)
\end{acks}



\bibliographystyle{ACM-Reference-Format}
\bibliography{chiea25-622}  


\begin{thebibliography}{33}


\ifx \showCODEN    \undefined \def \showCODEN     #1{\unskip}     \fi
\ifx \showDOI      \undefined \def \showDOI       #1{#1}\fi
\ifx \showISBNx    \undefined \def \showISBNx     #1{\unskip}     \fi
\ifx \showISBNxiii \undefined \def \showISBNxiii  #1{\unskip}     \fi
\ifx \showISSN     \undefined \def \showISSN      #1{\unskip}     \fi
\ifx \showLCCN     \undefined \def \showLCCN      #1{\unskip}     \fi
\ifx \shownote     \undefined \def \shownote      #1{#1}          \fi
\ifx \showarticletitle \undefined \def \showarticletitle #1{#1}   \fi
\ifx \showURL      \undefined \def \showURL       {\relax}        \fi
\providecommand\bibfield[2]{#2}
\providecommand\bibinfo[2]{#2}
\providecommand\natexlab[1]{#1}
\providecommand\showeprint[2][]{arXiv:#2}

\bibitem[Abbott(2020)]%
        {abbott2020cambridge}
\bibfield{author}{\bibinfo{person}{H~Porter Abbott}.} \bibinfo{year}{2020}\natexlab{}.
\newblock \bibinfo{booktitle}{\emph{The Cambridge introduction to narrative}}.
\newblock \bibinfo{publisher}{Cambridge University Press}.
\newblock


\bibitem[Ariyani(2023)]%
        {ariyani2023improving}
\bibfield{author}{\bibinfo{person}{Etika Ariyani}.} \bibinfo{year}{2023}\natexlab{}.
\newblock \showarticletitle{IMPROVING STUDENTS READING PROFICIENCY USING BIONIC METHOD (A CLASSROOM ACTION RESEARCH AT 10th GRADE STUDENTS)}.
\newblock \bibinfo{journal}{\emph{JOEL: Journal of Educational and Language Research}} \bibinfo{volume}{3}, \bibinfo{number}{5} (\bibinfo{year}{2023}), \bibinfo{pages}{215--226}.
\newblock


\bibitem[Association.({[n.\,d.]})]%
        {TP-toolbox-web}
\bibfield{author}{\bibinfo{person}{International~Dyslexia Association.}} \bibinfo{year}{[n.\,d.]}\natexlab{}.
\newblock \bibinfo{title}{Frequently Asked Questions About Dyslexia}.
\newblock \bibinfo{howpublished}{\url{http://www.interdys.org/}}.
\newblock


\bibitem[Berners-Lee et~al\mbox{.}(1994)]%
        {berners1994world}
\bibfield{author}{\bibinfo{person}{Tim Berners-Lee}, \bibinfo{person}{Robert Cailliau}, \bibinfo{person}{Ari Luotonen}, \bibinfo{person}{Henrik~Frystyk Nielsen}, {and} \bibinfo{person}{Arthur Secret}.} \bibinfo{year}{1994}\natexlab{}.
\newblock \showarticletitle{The world-wide web}.
\newblock \bibinfo{journal}{\emph{Commun. ACM}} \bibinfo{volume}{37}, \bibinfo{number}{8} (\bibinfo{year}{1994}), \bibinfo{pages}{76--82}.
\newblock


\bibitem[Berners-Lee and Cailliau(1990)]%
        {berners1990worldwideweb}
\bibfield{author}{\bibinfo{person}{Timothy~J Berners-Lee} {and} \bibinfo{person}{Robert Cailliau}.} \bibinfo{year}{1990}\natexlab{}.
\newblock \showarticletitle{WorldWideWeb: Proposal for a HyperText project}.
\newblock \bibinfo{journal}{\emph{World Wide Web Proposal}} (\bibinfo{year}{1990}).
\newblock


\bibitem[Blackpink(2024)]%
        {blackpink2024}
\bibfield{author}{\bibinfo{person}{Blackpink}.} \bibinfo{year}{2024}\natexlab{}.
\newblock \bibinfo{title}{{Blackpink}}.
\newblock \bibinfo{howpublished}{\url{https://en.wikipedia.org/wiki/Blackpink}}.
\newblock
\urldef\tempurl%
\url{https://en.wikipedia.org/wiki/Blackpink}
\showURL{%
\tempurl}
\newblock
\shownote{Accessed: 2024-09-07}.


\bibitem[Bommasani et~al\mbox{.}(2021)]%
        {bommasani2021opportunities}
\bibfield{author}{\bibinfo{person}{Rishi Bommasani}, \bibinfo{person}{Drew~A Hudson}, \bibinfo{person}{Ehsan Adeli}, \bibinfo{person}{Russ Altman}, \bibinfo{person}{Simran Arora}, \bibinfo{person}{Sydney von Arx}, \bibinfo{person}{Michael~S Bernstein}, \bibinfo{person}{Jeannette Bohg}, \bibinfo{person}{Antoine Bosselut}, \bibinfo{person}{Emma Brunskill}, {et~al\mbox{.}}} \bibinfo{year}{2021}\natexlab{}.
\newblock \showarticletitle{On the opportunities and risks of foundation models}.
\newblock \bibinfo{journal}{\emph{arXiv preprint arXiv:2108.07258}} (\bibinfo{year}{2021}).
\newblock


\bibitem[Brooke(1996)]%
        {brooke1996sus}
\bibfield{author}{\bibinfo{person}{John Brooke}.} \bibinfo{year}{1996}\natexlab{}.
\newblock \showarticletitle{Sus: a “quick and dirty’usability}.
\newblock \bibinfo{journal}{\emph{Usability evaluation in industry}} \bibinfo{volume}{189}, \bibinfo{number}{3} (\bibinfo{year}{1996}), \bibinfo{pages}{189--194}.
\newblock


\bibitem[Brysbaert(2019)]%
        {brysbaert2019many}
\bibfield{author}{\bibinfo{person}{Marc Brysbaert}.} \bibinfo{year}{2019}\natexlab{}.
\newblock \showarticletitle{How many words do we read per minute? A review and meta-analysis of reading rate}.
\newblock \bibinfo{journal}{\emph{Journal of memory and language}}  \bibinfo{volume}{109} (\bibinfo{year}{2019}), \bibinfo{pages}{104047}.
\newblock


\bibitem[Council({[n.\,d.]})]%
        {IELTS}
\bibfield{author}{\bibinfo{person}{British Council}.} \bibinfo{year}{[n.\,d.]}\natexlab{}.
\newblock \bibinfo{title}{IELTS}.
\newblock \bibinfo{howpublished}{\url{https://www.ielts.org/}}.
\newblock


\bibitem[Cullen et~al\mbox{.}(2013)]%
        {cullen2013effects}
\bibfield{author}{\bibinfo{person}{Jennifer Cullen}, \bibinfo{person}{Sue Keesey}, {and} \bibinfo{person}{Sheila~R Alber-Morgan}.} \bibinfo{year}{2013}\natexlab{}.
\newblock \showarticletitle{The effects of computer-assisted instruction using Kurzweil 3000 on sight word acquisition for students with mild disabilities}.
\newblock \bibinfo{journal}{\emph{Education and Treatment of Children}} (\bibinfo{year}{2013}), \bibinfo{pages}{87--103}.
\newblock


\bibitem[de~Santana et~al\mbox{.}(2013)]%
        {de2013firefixia}
\bibfield{author}{\bibinfo{person}{Vagner~Figueredo de Santana}, \bibinfo{person}{Rosimeire de Oliveira}, \bibinfo{person}{Leonelo Dell~Anhol Almeida}, {and} \bibinfo{person}{Marcia Ito}.} \bibinfo{year}{2013}\natexlab{}.
\newblock \showarticletitle{Firefixia: An accessibility web browser customization toolbar for people with dyslexia}. In \bibinfo{booktitle}{\emph{Proceedings of the 10th International Cross-Disciplinary Conference on Web Accessibility}}. \bibinfo{pages}{1--4}.
\newblock


\bibitem[Dickinson et~al\mbox{.}(2002)]%
        {dickinson2002ongoing}
\bibfield{author}{\bibinfo{person}{Anna Dickinson}, \bibinfo{person}{Peter Gregor}, {and} \bibinfo{person}{Alan~F Newell}.} \bibinfo{year}{2002}\natexlab{}.
\newblock \showarticletitle{Ongoing investigation of the ways in which some of the problems encountered by some dyslexics can be alleviated using computer techniques}. In \bibinfo{booktitle}{\emph{Proceedings of the fifth international ACM conference on Assistive technologies}}. \bibinfo{pages}{97--103}.
\newblock


\bibitem[Galuschka et~al\mbox{.}(2020)]%
        {galuschka2020effectiveness}
\bibfield{author}{\bibinfo{person}{Katharina Galuschka}, \bibinfo{person}{Ruth G{\"o}rgen}, \bibinfo{person}{Julia Kalmar}, \bibinfo{person}{Stefan Haberstroh}, \bibinfo{person}{Xenia Schmalz}, {and} \bibinfo{person}{Gerd Schulte-K{\"o}rne}.} \bibinfo{year}{2020}\natexlab{}.
\newblock \showarticletitle{Effectiveness of spelling interventions for learners with dyslexia: A meta-analysis and systematic review}.
\newblock \bibinfo{journal}{\emph{Educational Psychologist}} \bibinfo{volume}{55}, \bibinfo{number}{1} (\bibinfo{year}{2020}), \bibinfo{pages}{1--20}.
\newblock


\bibitem[Goodman et~al\mbox{.}(2022)]%
        {goodman2022lampost}
\bibfield{author}{\bibinfo{person}{Steven~M Goodman}, \bibinfo{person}{Erin Buehler}, \bibinfo{person}{Patrick Clary}, \bibinfo{person}{Andy Coenen}, \bibinfo{person}{Aaron Donsbach}, \bibinfo{person}{Tiffanie~N Horne}, \bibinfo{person}{Michal Lahav}, \bibinfo{person}{Robert MacDonald}, \bibinfo{person}{Rain~Breaw Michaels}, \bibinfo{person}{Ajit Narayanan}, {et~al\mbox{.}}} \bibinfo{year}{2022}\natexlab{}.
\newblock \showarticletitle{Lampost: Design and evaluation of an ai-assisted email writing prototype for adults with dyslexia}. In \bibinfo{booktitle}{\emph{Proceedings of the 24th International ACM SIGACCESS Conference on Computers and Accessibility}}. \bibinfo{pages}{1--18}.
\newblock


\bibitem[Laga et~al\mbox{.}(2006)]%
        {laga2006kurzweil}
\bibfield{author}{\bibinfo{person}{Kristen Laga}, \bibinfo{person}{Daniel Steere}, {and} \bibinfo{person}{Domenico Cavaiuolo}.} \bibinfo{year}{2006}\natexlab{}.
\newblock \showarticletitle{Kurzweil 3000}.
\newblock \bibinfo{journal}{\emph{Journal of Special Education Technology}} \bibinfo{volume}{21}, \bibinfo{number}{2} (\bibinfo{year}{2006}), \bibinfo{pages}{79}.
\newblock


\bibitem[Larco et~al\mbox{.}(2021)]%
        {larco2021moving}
\bibfield{author}{\bibinfo{person}{Andres Larco}, \bibinfo{person}{Jorge Carrillo}, \bibinfo{person}{Nelson Chicaiza}, \bibinfo{person}{Cesar Yanez}, {and} \bibinfo{person}{Sergio Luj{\'a}n-Mora}.} \bibinfo{year}{2021}\natexlab{}.
\newblock \showarticletitle{Moving beyond limitations: Designing the helpdys app for children with dyslexia in rural areas}.
\newblock \bibinfo{journal}{\emph{Sustainability}} \bibinfo{volume}{13}, \bibinfo{number}{13} (\bibinfo{year}{2021}), \bibinfo{pages}{7081}.
\newblock


\bibitem[Li et~al\mbox{.}(2019)]%
        {li2019impact}
\bibfield{author}{\bibinfo{person}{Qisheng Li}, \bibinfo{person}{Meredith~Ringel Morris}, \bibinfo{person}{Adam Fourney}, \bibinfo{person}{Kevin Larson}, {and} \bibinfo{person}{Katharina Reinecke}.} \bibinfo{year}{2019}\natexlab{}.
\newblock \showarticletitle{The impact of web browser reader views on reading speed and user experience}. In \bibinfo{booktitle}{\emph{Proceedings of the 2019 CHI conference on human factors in computing systems}}. \bibinfo{pages}{1--12}.
\newblock


\bibitem[McCarthy and Swierenga(2010)]%
        {mccarthy2010we}
\bibfield{author}{\bibinfo{person}{Jacob~E McCarthy} {and} \bibinfo{person}{Sarah~J Swierenga}.} \bibinfo{year}{2010}\natexlab{}.
\newblock \showarticletitle{What we know about dyslexia and web accessibility: a research review}.
\newblock \bibinfo{journal}{\emph{Universal Access in the Information Society}}  \bibinfo{volume}{9} (\bibinfo{year}{2010}), \bibinfo{pages}{147--152}.
\newblock


\bibitem[Morris et~al\mbox{.}(2018)]%
        {morris2018understanding}
\bibfield{author}{\bibinfo{person}{Meredith~Ringel Morris}, \bibinfo{person}{Adam Fourney}, \bibinfo{person}{Abdullah Ali}, {and} \bibinfo{person}{Laura Vonessen}.} \bibinfo{year}{2018}\natexlab{}.
\newblock \showarticletitle{Understanding the needs of searchers with dyslexia}. In \bibinfo{booktitle}{\emph{Proceedings of the 2018 CHI Conference on Human Factors in Computing Systems}}. \bibinfo{pages}{1--12}.
\newblock


\bibitem[Niklaus et~al\mbox{.}(2023)]%
        {niklaus2023digital}
\bibfield{author}{\bibinfo{person}{Aleena~Gertrudes Niklaus}, \bibinfo{person}{Tianyuan Cai}, \bibinfo{person}{Zoya Bylinskii}, {and} \bibinfo{person}{Shaun Wallace}.} \bibinfo{year}{2023}\natexlab{}.
\newblock \showarticletitle{Digital Reading Rulers: Evaluating Inclusively Designed Rulers for Readers With Dyslexia and Without}. In \bibinfo{booktitle}{\emph{Proceedings of the 2023 CHI Conference on Human Factors in Computing Systems}}. \bibinfo{pages}{1--17}.
\newblock


\bibitem[OpenAI({[n.\,d.]})]%
        {ChatGPT}
\bibfield{author}{\bibinfo{person}{OpenAI}.} \bibinfo{year}{[n.\,d.]}\natexlab{}.
\newblock \bibinfo{title}{ChatGPT}.
\newblock \bibinfo{howpublished}{\url{https://openai.com/}}.
\newblock


\bibitem[Ostiz-Blanco et~al\mbox{.}(2021)]%
        {ostiz2021improving}
\bibfield{author}{\bibinfo{person}{Mikel Ostiz-Blanco}, \bibinfo{person}{Javier Bernacer}, \bibinfo{person}{Irati Garcia-Arbizu}, \bibinfo{person}{Patricia Diaz-Sanchez}, \bibinfo{person}{Luz Rello}, \bibinfo{person}{Marie Lallier}, {and} \bibinfo{person}{Gonzalo Arrondo}.} \bibinfo{year}{2021}\natexlab{}.
\newblock \showarticletitle{Improving reading through videogames and digital apps: A systematic review}.
\newblock \bibinfo{journal}{\emph{Frontiers in psychology}}  \bibinfo{volume}{12} (\bibinfo{year}{2021}), \bibinfo{pages}{652948}.
\newblock


\bibitem[{Prolific}(2023)]%
        {prolific}
\bibfield{author}{\bibinfo{person}{{Prolific}}.} \bibinfo{year}{2023}\natexlab{}.
\newblock \bibinfo{title}{Prolific - Online Participant Recruitment for Surveys and Market Research}.
\newblock
\newblock
\urldef\tempurl%
\url{https://www.prolific.com/}
\showURL{%
\tempurl}


\bibitem[Reading({[n.\,d.]})]%
        {BionicReading}
\bibfield{author}{\bibinfo{person}{Bionic Reading}.} \bibinfo{year}{[n.\,d.]}\natexlab{}.
\newblock \bibinfo{title}{Bionic Reading}.
\newblock \bibinfo{howpublished}{\url{https://bionic-reading.com/}}.
\newblock


\bibitem[Rello and Baeza-Yates(2013)]%
        {rello2013good}
\bibfield{author}{\bibinfo{person}{Luz Rello} {and} \bibinfo{person}{Ricardo Baeza-Yates}.} \bibinfo{year}{2013}\natexlab{}.
\newblock \showarticletitle{Good fonts for dyslexia}. In \bibinfo{booktitle}{\emph{Proceedings of the 15th international ACM SIGACCESS conference on computers and accessibility}}. \bibinfo{pages}{1--8}.
\newblock


\bibitem[Rello and Baeza-Yates(2014)]%
        {rello2014evaluation}
\bibfield{author}{\bibinfo{person}{Luz Rello} {and} \bibinfo{person}{Ricardo Baeza-Yates}.} \bibinfo{year}{2014}\natexlab{}.
\newblock \showarticletitle{Evaluation of DysWebxia: a reading app designed for people with dyslexia}. In \bibinfo{booktitle}{\emph{Proceedings of the 11th Web for All Conference}}. \bibinfo{pages}{1--10}.
\newblock


\bibitem[Rello et~al\mbox{.}(2014)]%
        {rello2014computer}
\bibfield{author}{\bibinfo{person}{Luz Rello}, \bibinfo{person}{Clara Bayarri}, \bibinfo{person}{Yolanda Otal}, {and} \bibinfo{person}{Martin Pielot}.} \bibinfo{year}{2014}\natexlab{}.
\newblock \showarticletitle{A computer-based method to improve the spelling of children with dyslexia}. In \bibinfo{booktitle}{\emph{Proceedings of the 16th international ACM SIGACCESS conference on Computers \& accessibility}}. \bibinfo{pages}{153--160}.
\newblock


\bibitem[Rello et~al\mbox{.}(2012)]%
        {rello2012layout}
\bibfield{author}{\bibinfo{person}{Luz Rello}, \bibinfo{person}{Gaurang Kanvinde}, {and} \bibinfo{person}{Ricardo Baeza-Yates}.} \bibinfo{year}{2012}\natexlab{}.
\newblock \showarticletitle{Layout guidelines for web text and a web service to improve accessibility for dyslexics}. In \bibinfo{booktitle}{\emph{Proceedings of the international cross-disciplinary conference on web accessibility}}. \bibinfo{pages}{1--9}.
\newblock


\bibitem[Snell(2024)]%
        {snell2024no}
\bibfield{author}{\bibinfo{person}{Joshua Snell}.} \bibinfo{year}{2024}\natexlab{}.
\newblock \showarticletitle{No, Bionic Reading does not work}.
\newblock \bibinfo{journal}{\emph{Acta Psychologica}}  \bibinfo{volume}{247} (\bibinfo{year}{2024}), \bibinfo{pages}{104304}.
\newblock


\bibitem[Stodden et~al\mbox{.}(2012)]%
        {stodden2012use}
\bibfield{author}{\bibinfo{person}{Robert~A Stodden}, \bibinfo{person}{Kelly~D Roberts}, \bibinfo{person}{Kiriko Takahashi}, \bibinfo{person}{Hye~Jin Park}, {and} \bibinfo{person}{Norma~Jean Stodden}.} \bibinfo{year}{2012}\natexlab{}.
\newblock \showarticletitle{Use of text-to-speech software to improve reading skills of high school struggling readers}.
\newblock \bibinfo{journal}{\emph{Procedia Computer Science}}  \bibinfo{volume}{14} (\bibinfo{year}{2012}), \bibinfo{pages}{359--362}.
\newblock


\bibitem[{Tate}(nd)]%
        {sargentCarnation}
\bibfield{author}{\bibinfo{person}{{Tate}}.} \bibinfo{year}{n.d.}\natexlab{}.
\newblock \bibinfo{title}{{Carnation, Lily, Lily, Rose by John Singer Sargent}}.
\newblock \bibinfo{howpublished}{\url{https://www.tate.org.uk/art/artworks/sargent-carnation-lily-lily-rose-n01615}}.
\newblock
\urldef\tempurl%
\url{https://www.tate.org.uk/art/artworks/sargent-carnation-lily-lily-rose-n01615}
\showURL{%
\tempurl}
\newblock
\shownote{Accessed: 2024-09-11}.


\bibitem[Topac(2012)]%
        {topac2012development}
\bibfield{author}{\bibinfo{person}{V Topac}.} \bibinfo{year}{2012}\natexlab{}.
\newblock \showarticletitle{The development of a text customization tool for existing web sites}. In \bibinfo{booktitle}{\emph{Text Customization for Readability Symposium}}.
\newblock


\end{thebibliography}

%
\appendix
\section{Experiment Details}
\label{detail}

\subsection{Bionic Reading}
\label{bionic}
Bionic Reading is an application that present the first one or few character with bold effect. The Example of Bionic Reading is shown in Figure \ref{Bionic Reading}. 

\subsection{Bionic Reading Data}
\label{Bionicreading}
\begin{center}
\begin{tabular}{|m{8cm}|} %
\hline
\textbf{An Example of Bionic Reading} \\ \hline
\textbf{Black}Pink \textbf{i}s \textbf{a} \textbf{popu}lar \textbf{Sou}th \textbf{Kor}ean \textbf{gi}rl \textbf{gro}up \textbf{consi}sting \textbf{o}f members \textbf{Jisoo}, \textbf{Jenni}e, Rosé, and \textbf{Li}sa. \textbf{Th}ey \textbf{a}re \textbf{kno}wn \textbf{f}or \textbf{the}ir \textbf{energ}etic \textbf{performanc}es, diverse \textbf{mus}ic \textbf{style}s, and \textbf{fashio}nable \textbf{ima}ge. \textbf{Wi}th \textbf{hi}ts \textbf{li}ke "\textbf{D}DU-\textbf{D}U \textbf{D}DU-\textbf{D}U,“ \textbf{a}nd "\textbf{H}ow \textbf{Y}ou \textbf{Li}ke \textbf{Th}at," \textbf{Black}Pink \textbf{h}as \textbf{gai}ned \textbf{glo}bal \textbf{recogn}ition \textbf{a}nd \textbf{a} \textbf{str}ong \textbf{f}an \textbf{follo}wing.\\ \hline
\end{tabular}
\end{center}

\subsection{Attention Check}
\label{attention}
In our attention check, participants are asked to answer the question where including the instruction that the correct answer is "Water". Participants who fail in this question will be marked as not focused.

\begin{tabular}{|m{8cm}|}

\hline
In the modern era, explorers and archaeologists uncovered the secrets of the pyramid's chambers. The stories of Djoser, Imhotep, and the countless hands that had shaped the monument were revealed, shedding light on the ancient world's mysteries. \textbf{To show that you have read the instructions carefully, please ignore the items below about the explorers' findings and instead choose "Water".} Based on the information in the preceding paragraph, which of these objects did explorers find? \\ \hline
\begin{tabbing}
o \hspace{0.2cm} Gold \\
o \hspace{0.2cm} Diamond \\
o \hspace{0.2cm} Rosewood \\
o \hspace{0.2cm} Water \\
o \hspace{0.2cm} Stele \\
\end{tabbing} \\ \hline
\end{tabular}

\subsection{Dyslexia Checklist}
\label{Dyslexia Checklist}
We use the Dyslexia checklist \ref{table:dyslexia_checklist} to ask participants to evaluate their extent of different difficulties in reading.

\begin{table*}[hhbp]
\centering
\caption{\textbf{Dyslexia Checklist for the Experiment}}
\renewcommand{\arraystretch}{1} 
\begin{tabular*}{\textwidth}{@{\extracolsep{\fill}} m{0.2\linewidth} m{0.1\linewidth} m{0.6\linewidth}}
\hline
\textbf{Term} & \textbf{Scale} & \textbf{Description} \\[0.5em] 
\hline
Understanding & 1--7 & To what extent do you have difficulty understanding the meaning of sentences or paragraphs, even if individual words can be recognized? \\[0.1em]\\
Recognition   & 1--7 & To what extent do you struggle to correctly and fluently recognize letters and words, which can lead to slow reading speed and misinterpretation of words? \\[0.1em]\\
Memory        & 1--7 & To what extent do you struggle to remember what has been read, especially understanding longer texts or story plots? \\[0.1em]\\
Decoding      & 1--7 & To what extent do you have difficulty blending letters into words and understanding word pronunciation rules, affecting reading fluency and comprehension? \\[0.1em]\\
Attention     & 1--7 & To what extent do you have difficulty maintaining focus while reading for an extended period, leading to easy distractions? \\[0.1em]\\
Visual Disturbance & 1--7 & How frequently do you encounter visual disturbances during reading, such as letters or words appearing distorted, jumbled, or overlapping? \\
\hline
\end{tabular*}
\label{table:dyslexia_checklist}
\end{table*}

\subsection{Reasons Using Prolific}
Prolific encourages participants to disclose any health-related conditions, including dyslexia, allowing researchers to recruit specific individuals with relevant health conditions. Second, to control for factors such as time of day and time zone, which could potentially impact participants’ cognitive function, Prolific allows us to limit recruitment to participants within the same time zone.

\subsection{Recall Question}
\label{Recall Question}
Two examples of our recall questions are given below:

\begin{tabular}{|m{8cm}|}
\hline
\textbf{Where is the Step Pyramid of Djoser at?} \\ \hline
o \hspace{0.2cm} Saqquira  \\
o \hspace{0.2cm} Saqqara  \\
o \hspace{0.2cm} Saqqura  \\
o \hspace{0.2cm} Saqqarua  \\ \hline
\end{tabular}

\begin{tabular}{|m{8cm}|}
\hline
\textbf{Which king in ancient Egypt does this article discuss?} \\ \hline
Please input your answer: \hspace{0.2cm} \\ \hline
\end{tabular}

\subsection{Retrieve Question}
\label{Retrieve Question}
An example of our retrieve question is given below:

\begin{tabular}{|m{8cm}|}
\hline
\textbf{Please retrieve the article and provide as many details as possible (such as what specific data the article presents, what names appear, and the relationships between the characters and events, etc.)} \\ \hline
Please input your answer: \hspace{0.2cm} \\ \hline
\end{tabular}

\subsection{Default Prompt for GPT-4}
\label{default prompt}
This prompt is used for our experiment, as well as the default prompt in our software demo (default model). It is designed to be used in general situations but not for particular articles. The detailed prompt is displayed below:

\begin{tabular}{|m{8cm}|}
\hline
\textbf{You are an intelligent reader helper and you will be given a string of text in string format, please annotate it by adding tags following these instructions:} \\ \hline

1. Please annotate every date, number, location, and name of people or events in the paragraph by adding \texttt{<strong>} tags around them. \\

2. Please highlight sentences and phrases in the paragraph that can summarize the core content of the paragraph or serve as a conclusion to the description by adding \texttt{<mark>} tags around them. \\

3. Please underline sentences and phrases in the paragraph that are unusual or need to be particularly noted by adding \texttt{<u>} tags around them. \\

4. You can add as many \texttt{<mark>}, \texttt{<strong>}, or \texttt{<u>} tags in one paragraph as necessary to highlight or bold important text. \\

5. Please make sure to use and only use the 3 types of annotations above to annotate each paragraph of the text. \\

6. Don't make the highlights or underlines too long or too often if it is not necessary. \\

7. You are allowed to add only the above previously mentioned HTML tags, and that's the only change you can make to the text. YOUR OUTPUT MUST KEEP THE CONTENT OF THE ARTICLE THE SAME AS THE ORIGINAL ONE. \\

8. Your output should only contain the marked text with added tags, which can be directly presented in HTML. Don’t add anything else like "Here is your output" and so on. \\

9. Keep the original language; i.e., if the context was given in Chinese, your output should be Chinese as well. \\

\hline
\end{tabular}

\subsection{GPT-4 Evaluation Criteria}
\label{Criteria}
Here is the prompt for GPT-4 to give the score, we use a one-shot learning method to give GPT an example of a 6-point answer:
\\
\begin{tabular}{|m{8cm}|}
\hline
Please play the role of a rater and help me rate some answers. you will be given an article. Please read it, and you will be given some information about this article. I need you to score each item by their completeness and accuracy from 0 to 10.
\\A 0-point represents the entrance is very poor and basically contains no correct or important information and a 10 means the entrance is almost perfect.
\\A 5-point answer should have some details correct but misses or get some key information wrong, and the overall understanding of the article is partially correct.
\\A 7-point entrance should contain some correct details, such as the correct name, time, data, etc., or provide a not-bad summary of the overall article. However, it may be a lack of coherent logic or could miss some important information.
\\A 9-point entrance should contain most of the correct details, such as the correct name, time, data, etc., and it should also contain a logically coherent and accurate summary of the full text.
\\
\\
Here is the original article\\
*****\\
ORIGINAL ARTICLE\\
*****
\\
Now you should directly give a score and the reason you give that score, and here is an example of 6-point entrance:
The entrance is: 10.5 m high, with 13 false doors, there were tombs made of mud and clay before stone pyramids, the third Egyptian dynasty was the first to build of stone.
\\
And the answer is: 
\\
Score: 6\\
The entrance provides important details such as the height of the wall (10.5 meters) and the number of false doors (13). It also correctly mentions that tombs were made of mud and clay before the construction of stone pyramids and that the Third Dynasty of Egypt was the first to build with stone. However, it could have provided more information about the Step Pyramid itself, such as its final dimensions or its significance in Egyptian history. And its logic is not very coherent.
\\ \hline
\end{tabular}

\subsection{Subjective Evaluation}
\label{sub_1}
The subjective evaluation consists of a system usability scale (see Table \ref{tab:sbj_eva}) and a general subjective evaluation scale (see Table \ref{tab:sbj_eva_2}).

\begin{table*}[!htbp]
\centering
\caption{\textbf{Subjective Evaluation - System Usability Scales}}
\label{tab:sbj_eva}
\resizebox{\textwidth}{!}{%
\begin{tabular}{p{4cm}p{4cm}p{1cm}p{1cm}p{1cm}p{1.2cm}}
\hline
\multicolumn{2}{c}{\textbf{System Usability Scales}} & \multicolumn{2}{c}{\textbf{Mean (SD)}} & \multicolumn{1}{c}{\multirow{2}{*}{\textbf{\makecell{Statistics\\(F(1, 95))}}}} & \multicolumn{1}{c}{\multirow{2}{*}{\textbf{p-value}}} \\ \cline{1-4}
\multicolumn{1}{c}{\textbf{Conventional}} & \multicolumn{1}{c}{\textbf{LARF}} & \multicolumn{1}{c}{\textbf{Conventional}} & \multicolumn{1}{c}{\textbf{LARF}} &  &  \\ \hline
I believe that I would frequently like to read articles with these types of bold labels on certain occasions. & I believe that I would frequently like to read articles with these types of highlights, underlines, or bold labels on certain occasions. & 3.35 (2.07) & 3.77 (1.96) & 1.073 & p = .303 \\ \hline
I think understanding these bold labels was not difficult for me. & I think understanding these highlights, underlines, or bold labels was not difficult for me. & 3.96 (1.78) & 4.31 (1.84) & .927 & p = .338 \\ \hline
I believe I would need the support of a technical person to read an article with these bold labels.[reversed-scale] & I believe I would need the support of a technical person to read an article with these highlights, underlines, or bold labels.[reversed-scale] & 5.55 (1.62) & 5.29 (1.86) & .538 & p = .465 \\ \hline
I found that the bold labels were well-integrated. & I found that the highlights, underlines, or bold labels were well-integrated. & 3.63 (2.02) & 4.23 (1.68) & 2.500 & p = .117 \\ \hline
I would imagine that most people would learn to read with these bold labels very quickly. & I would imagine that most people would learn to read with these highlights, underlines, or bold labels very quickly. & 3.96 (1.84) & 4.65 (1.89) & 2.543 & p = .114 \\ \hline
I felt very confident reading with the bold labels. & I felt very confident reading with the highlights, underlines, or bold labels. & 4.06 (1.73) & 4.40 (1.83) & .859 & p = .356 \\ \hline
\multicolumn{6}{l}{\begin{tabular}[c]{@{}l@{}}Notes: \\ (1) Standard errors are in parentheses; \\ (2) *p \textless{} 0.1, **p \textless{} 0.05, ***p \textless{} 0.01 \\ (3) SUS-3 is a reversed-scale question\end{tabular}} \\
\end{tabular}%
}
\end{table*}

\begin{table*}[!htbp]
\centering
\caption{\textbf{Subjective Evaluation - All}}
\label{tab:sbj_eva_2}
\begin{tabular*}{\textwidth}{@{\extracolsep{\fill}}p{2.5cm} p{6cm} p{2cm} p{2cm} p{1.5cm} p{1.5cm}@{}}
\hline
\multirow{2}{*}{\textbf{Metrics}} & \multirow{2}{*}{\textbf{Question}} & \multicolumn{2}{c}{\textbf{Mean (SD)}} & \multirow{2}{*}{\textbf{\makecell{Statistics\\(F(1, 95))}}} & \multirow{2}{*}{\textbf{p-value}} \\ \cline{3-4}
                                  &                                    & \textbf{Conventional}  & \textbf{LARF} &                                      &                                   \\ \hline
\textbf{Satisfaction}             & What is your overall satisfaction with this kind of presentation (highlights, underlines, or bold labels annotations/bold labels annotations) when you read articles?        & 3.76 (1.92)            & 4.42 (1.84)   & 2.994                      & .087*                           \\ \hline
\textbf{Helpfulness}              & To what extent do you think you will continue to use this kind of presentation (highlights, underlines, or bold labels annotations/bold labels annotations) in future reading?             & 3.14 (1.76)            & 4.29 (1.99)   & 9.104                       & .003**                          \\ \hline
\textbf{\makecell[tl]{Intention for\\ Future Use}} & To what extent do you believe the marks in the articles helped you concentrate on the key information? & 2.94 (1.89)            & 3.92 (2.01)   & 6.111                     & .015*                           \\ \hline
\textbf{Recommendation}           & To what extent will you recommend this kind of presentation (highlights, underlines, or bold labels annotations/bold labels annotations) to others?    & 3.18 (1.87)            & 4.42 (1.97)   & 10.034                     & .002**                           \\ \hline
\textbf{\makecell[tl]{Intention for\\ Widespread Usage}} & Do you think this kind of presentation (highlights, underlines, or bold labels annotations/bold labels annotations) is suitable for widespread use in other contexts? For example, in special exam papers for people with reading disabilities, integrated into e-readers, or for online academic paper reading? & 3.69 (2.10)            & 4.96 (1.96)   & 9.388                     & .003**                          \\ \hline
\multicolumn{2}{l}{\begin{tabular}[c]{@{}l@{}}Notes: \\ (1) Standard errors are in parentheses; \\ (2) *p \textless 0.1, **p \textless 0.05, ***p \textless 0.01\end{tabular}} \\ \hline
\end{tabular*}
\end{table*}

\subsection{Participants}
\label{participants}
Participants' demographics in the experiment can be found in Table \ref{demographic}.

\begin{table*}[htbp]
\centering
\caption{\textbf{Participants Demographic in the Experiment}}
\label{demographic}
\begin{tabular*}{\textwidth}{@{\extracolsep{\fill}} ll|lr|lr}
\hline
\multicolumn{2}{c|}{Gender}                 & \multicolumn{2}{c|}{Age} & \multicolumn{2}{c}{Education} \\ \hline
Male               & 95    & 18-24         & 11   & Less than high school                          & 2    \\
Female             & 50    & 25-34         & 56   & High School graduate                           & 58   \\
Non-binary/Unknown & 5     & 35-44         & 44   & Bachelor degree (or currently in processing)   & 56   \\
                   &       & 45-54         & 26   & Master degree (or currently in processing)     & 26   \\
                   &       & 55-64         & 12   & Doctor degree (or currently in processing)     & 6    \\
                   &       & 65-74         & 0    &                                                &      \\
                   &       & 75+           & 0    &                                                &      \\ \hline
\end{tabular*}
\end{table*}

\section{Software Demo}
\label{software demo}
\begin{figure*}[htbp]
  \centering
  \includegraphics[width=0.8\textwidth]{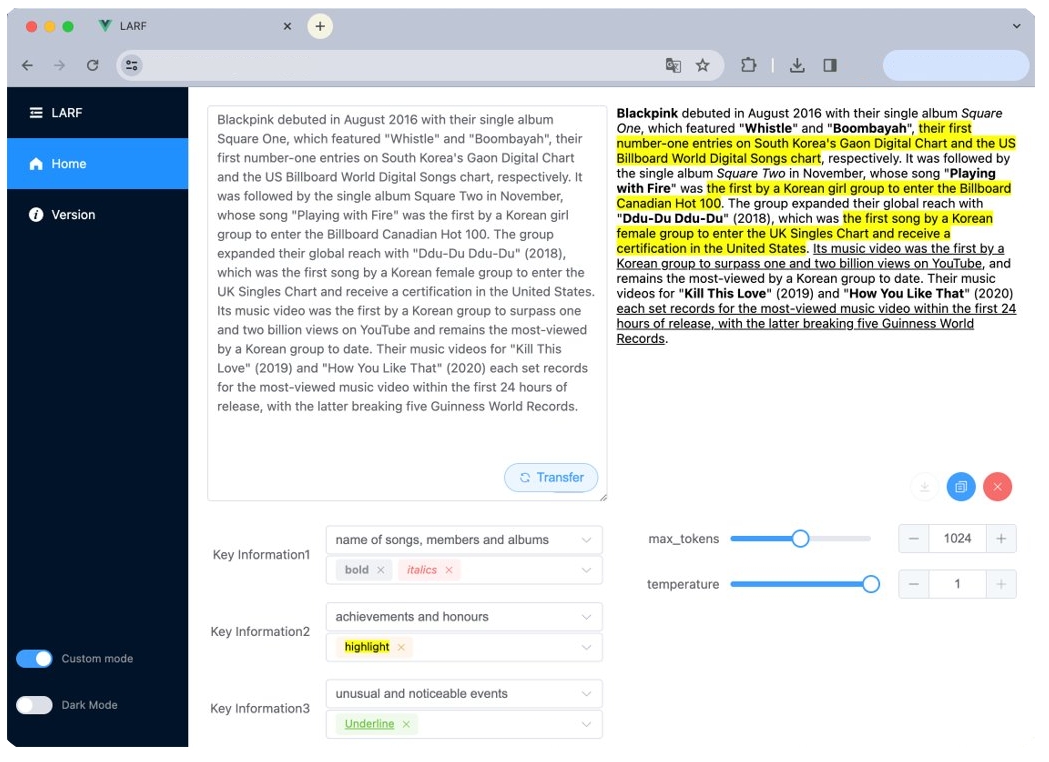}
  \Description{This screenshot shows a web interface with a left sidebar labeled "LARF," containing navigation items such as "Home" and "Version." The main panel displays a text passage about the K-pop group Blackpink, including details on their debut, chart performance, and global reach. Portions of the text are highlighted in yellow, indicating specific annotations or emphasis. Below the text, there is a button labeled "Transfer. " On the left side, there are a series of labeled fields, including "Key Information," "abilities," "achievements and honours," "unusual and noticeable events," and "others," each accompanied by text boxes. On the right side of these fields, there are numeric input boxes or sliders labeled "max_tokens" and "temperature," suggesting adjustable parameters for a language model or text-processing feature. Overall, the interface allows users to view and annotate a passage of text, categorize its content (e.g., key information or achievements), and potentially configure generation or processing settings (e.g., max_tokens, temperature). The browser window at the top shows typical navigation controls and an address bar, indicating that this is a web-based application.}
  \caption{The demo of the custom mode of LARF software application on PC.}
  \label{fig: demo}
\end{figure*}

\begin{figure*}[htbp]
  \centering
  \includegraphics[width=1\textwidth]{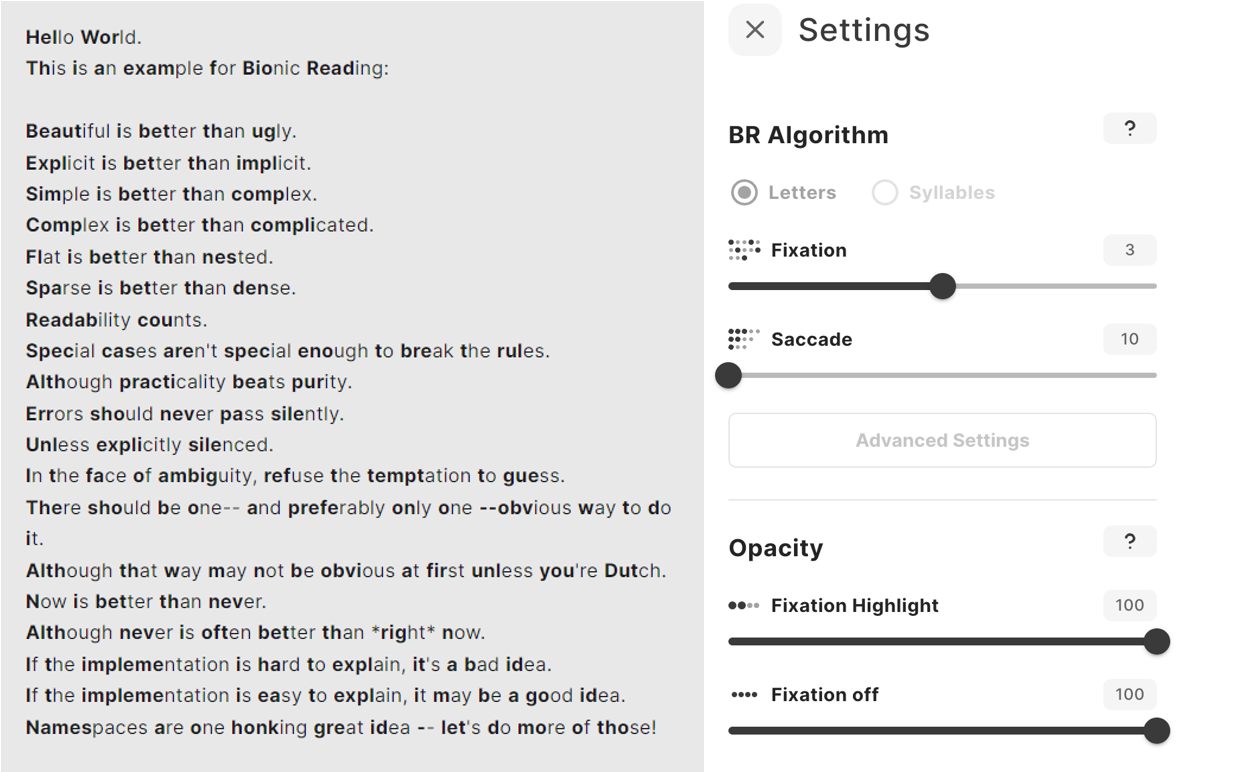}
  \Description{This screenshot is split into two main sections. On the left, there is a text area showing a sample passage that demonstrates Bionic Reading techniques. The text includes short statements such as “Hello World. This is an example for Bionic Reading,” followed by lines like “Beautiful is better than ugly,” “Explicit is better than implicit,” and so forth, reflecting stylistic or readability guidelines.

On the right, there is a vertical “Settings” panel labeled “BR Algorithm.” Two toggle options—“Letters” and “Syllables”—appear at the top, indicating different methods for applying the Bionic Reading algorithm. Below them, there are sliders labeled “Fixation” and “Saccade,” each adjustable from 0 to 100, allowing users to customize how text is bolded or spaced to guide the eye. An “Advanced Settings” button is also present.

Farther down, additional sliders are shown under the heading “Opacity,” including “Fixation Highlight” and “Fixation off,” each with numeric ranges, which control the transparency and visibility of bolded segments or other text effects. Overall, the left side displays real-time changes to the reading text as the user adjusts the parameters on the right, enabling a customized reading experience.}

  \caption{An example of the result and parameters of Bionic Reading}
  \label{Bionic Reading}
\end{figure*}

The LARF software demo shown in Figure \ref{fig: demo} is an interactive interface that users can open in a browser via a link. Users can copy and paste the text into the text box on the left, and by clicking the "Transfer" button, they can obtain the annotated text in the text box on the right. By checking the "Custom mode" option on the left, users can activate the custom prompt feature. When Custom mode is off, LARF will process the text using the same prompt as in the previous experiments. When Custom mode is on, users can enter the information they want to be annotated (such as the names of songs, members, and albums shown in the figure) and specify how they want this information to be annotated in the Key Information section below.

\section{Supplemental Figures}

\label{Supplemental Figures}
\subsection{Subjective Evaluation Result in Experiment}
\label{Subjective Evaluation Result in Experiment 1}
The subjective evaluation includes system usability ($M_{conventional} = 4.09, SD = 1.42; M_{LARF} = 4.43, SD = 1.36; F(1,95) = 1.469, p = .229$), satisfaction of the tool ($M_{conventional}=3.76, SD = 1.92; M_{LARF} = 4.42, SD = 1.84; F(1,95) = 2.994, p = .087$), perceived helpfulness ($M_{conventional} = 3.14, SD = 1.76; M_{LARF} =4.29, SD = 1.99; F(1,95) = 9.104, p = .003$), intention for future usage ($M_{conventional}=2.94, SD = 1.89;  M_{LARF}=3.92, SD = 2.01; F(1,95) = 6.111, p = .015$), recommend ($M_{conventional}=3.18, SD = 1.87;  M_{LARF}=4.42, SD = 1.97; F(1,95) = 10.034, p =.002$), and widespread usage ($M_conventional=3.69, SD = 2.10; M_{LARF} =4.96, SD = 1.96; F(1,95) = 9.388, p =.003$). Participants in the LARF condition expressed a favourable inclination towards customizing the LARF tool. This preference was quantitatively reflected, with the mean score for the desire to customize LARF being 5.04 (SD = 1.41). 

\begin{figure*}[!htbp]
  \centering
  \includegraphics[width=1\textwidth]{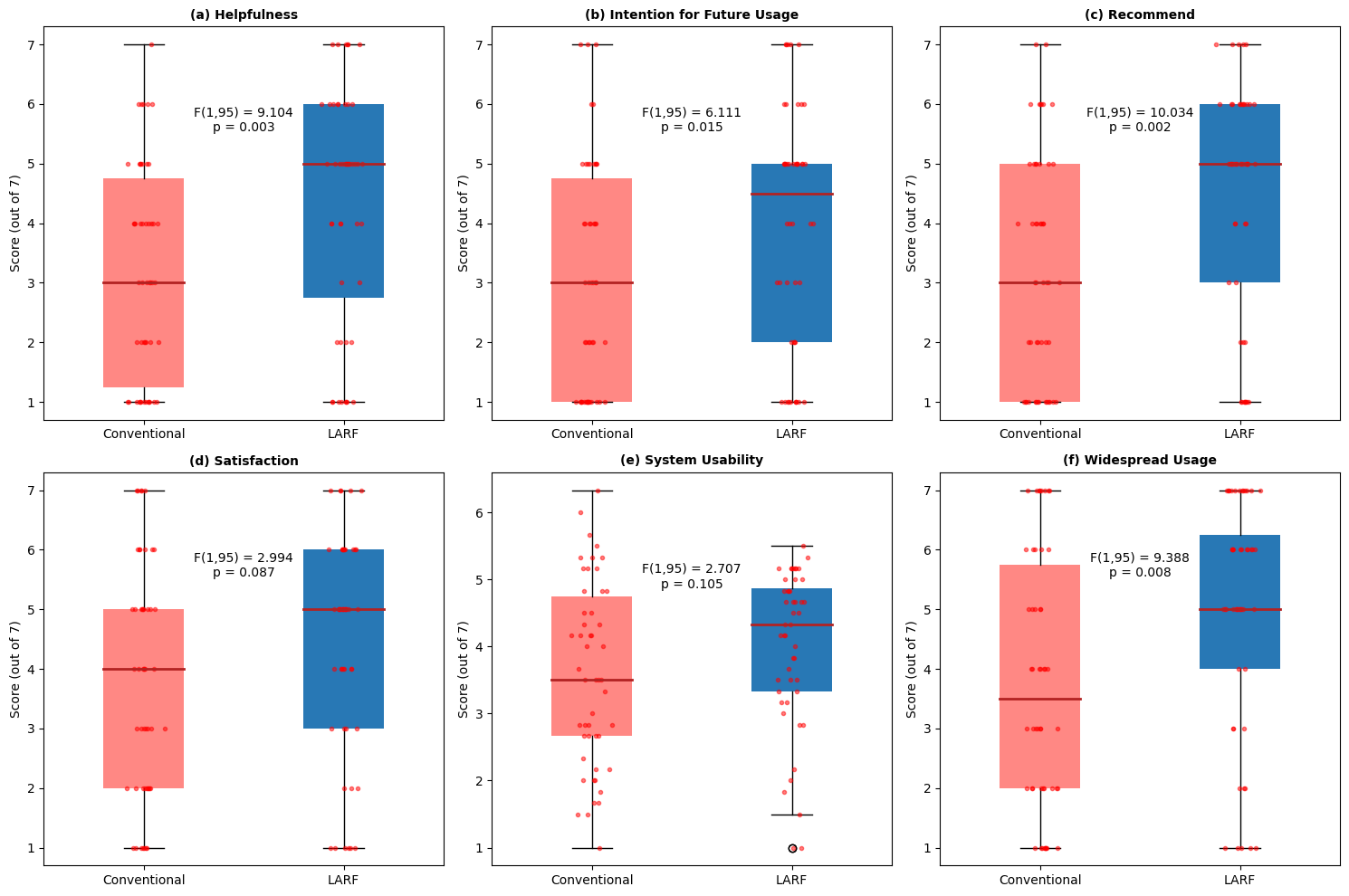}
  \Description{This figure consists of six boxplots arranged in a 2×3 grid, labeled (a) through (f). Each subplot compares two groups, Conventional (shown in red on the left) and LARF (shown in blue on the right), with scores on the y-axis. The subplots are:
  
(a) Helpfulness: Scores on a 1–7 scale, where higher indicates more perceived helpfulness.
(b) Intention for Future Usage: Scores on a 1–7 scale, representing how likely participants are to continue using the annotation method.
(c) Recommend: Scores on a 1–7 scale, showing how likely participants are to recommend this annotation method.
(d) Satisfaction: Scores on a 1–7 scale, reflecting overall satisfaction with the annotation style.
(e) System Usability: Scores on a 1–7 scale, assessing usability aspects of the method.
(f) Widespread Usage: Scores on a 1–7 scale, indicating whether participants believe the method is suitable for broader application.

Each boxplot shows the median as a horizontal line within the box (representing the interquartile range), whiskers extending to data within 1.5 times the interquartile range, and outlier points as circles beyond the whiskers. Statistical results (F-values and p-values) are displayed above each boxplot. Overall, the LARF group (blue) generally appears to score higher than the Conventional group (red) across these six subjective evaluation metrics.}

  \caption{The subjective evaluation result. Participants with dyslexia exhibited a clear preference for LARF, considering text annotated with LARF to be effective, user-friendly, and worthy of broader adoption in various contexts. }
  \label{fig: Fig. 10}
\end{figure*}

\subsection{Post Hoc Evaluation}
The Post hoc evaluation for the experiment: Given that individuals with dyslexia may encounter varying types and degrees of reading challenges, we categorized each symptom in the dyslexia checklist into "severe" and "mild". The red line depicted in the figure represents the performance of users facing more significant challenges in that specific item. The plot shows that LARF significantly improved recall, retrieval, and comprehension performance in individuals with more severe symptoms.

\begin{figure*}[!htbp]
  \centering
  \includegraphics[width=0.8\textwidth]{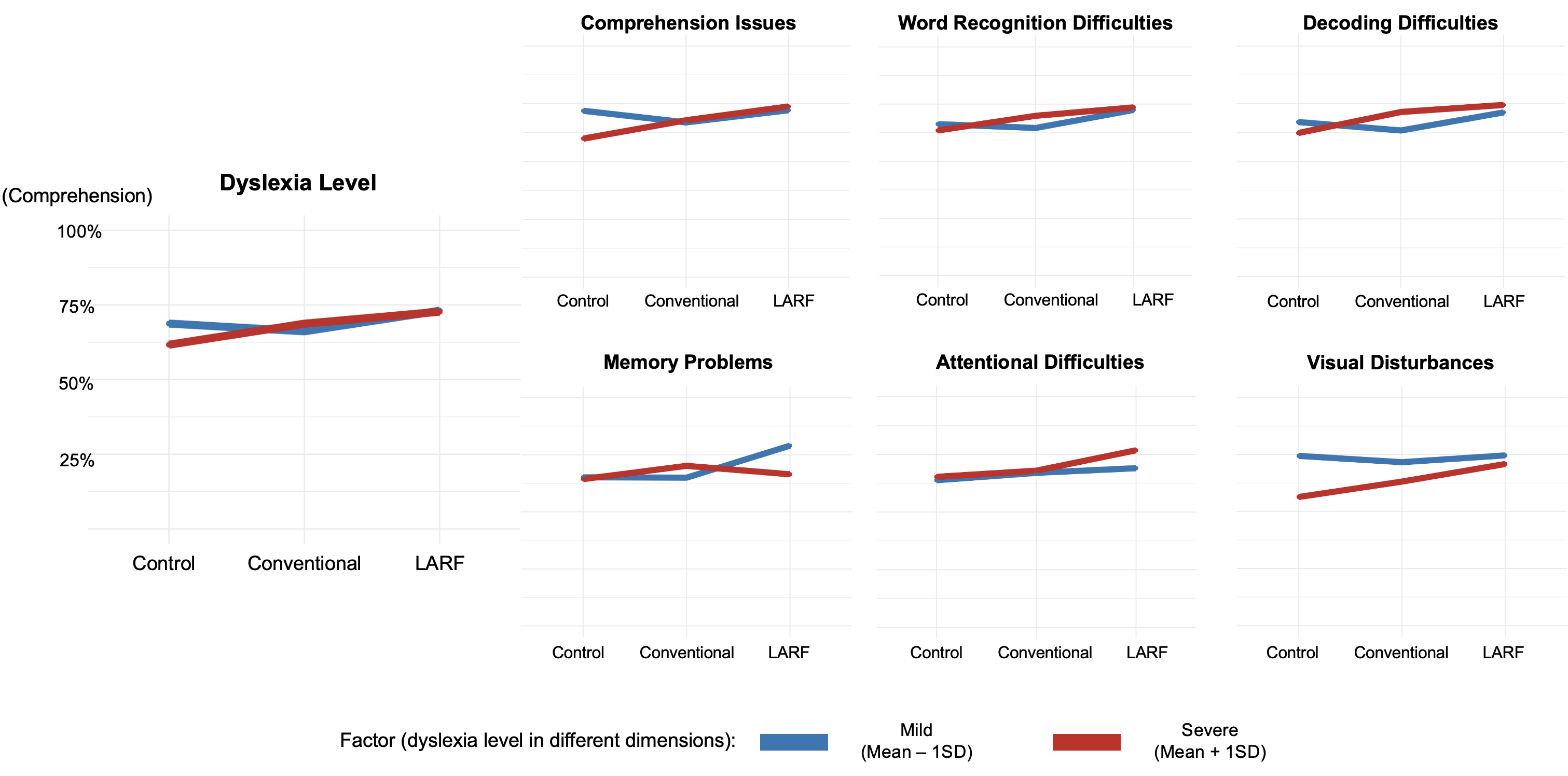}
  \Description{This figure consists of six small line charts arranged in a 2×3 grid, each labeled with a different dyslexia-related dimension: “Comprehension Issues,” “Word Recognition Difficulties,” “Decoding Difficulties,” “Memory Problems,” “Attentional Difficulties,” and “Visual Disturbances.” In each chart, the x-axis shows three groups (Control, Conventional, and LARF), and the y-axis (ranging from 0\% to 100\%) represents the severity or level for that particular dimension.

Each chart has two lines: one for participants with mild dyslexia (in blue) and one for participants with severe dyslexia (in red). The legend at the bottom indicates “Mild (Mean ± SD)” in blue and “Severe (Mean ± SD)” in red. The lines connect the mean values across the three groups, with vertical markers suggesting standard deviations. Overall, these six plots illustrate how mild and severe dyslexia levels differ in each dimension (e.g., comprehension, memory, attention) under the three different reading approaches (Control, Conventional, LARF).}

  \caption{Post hoc evaluation for comprehension performance. The y-axis represents the accuracy of reading comprehension. In the group with severe symptoms, LARF exhibited significant improvement compare to the conventional group and control group.}
  \label{fig:subfig comprehension}
\end{figure*}

\begin{figure*}[!htbp]
  \centering
  \includegraphics[width=0.8\textwidth]{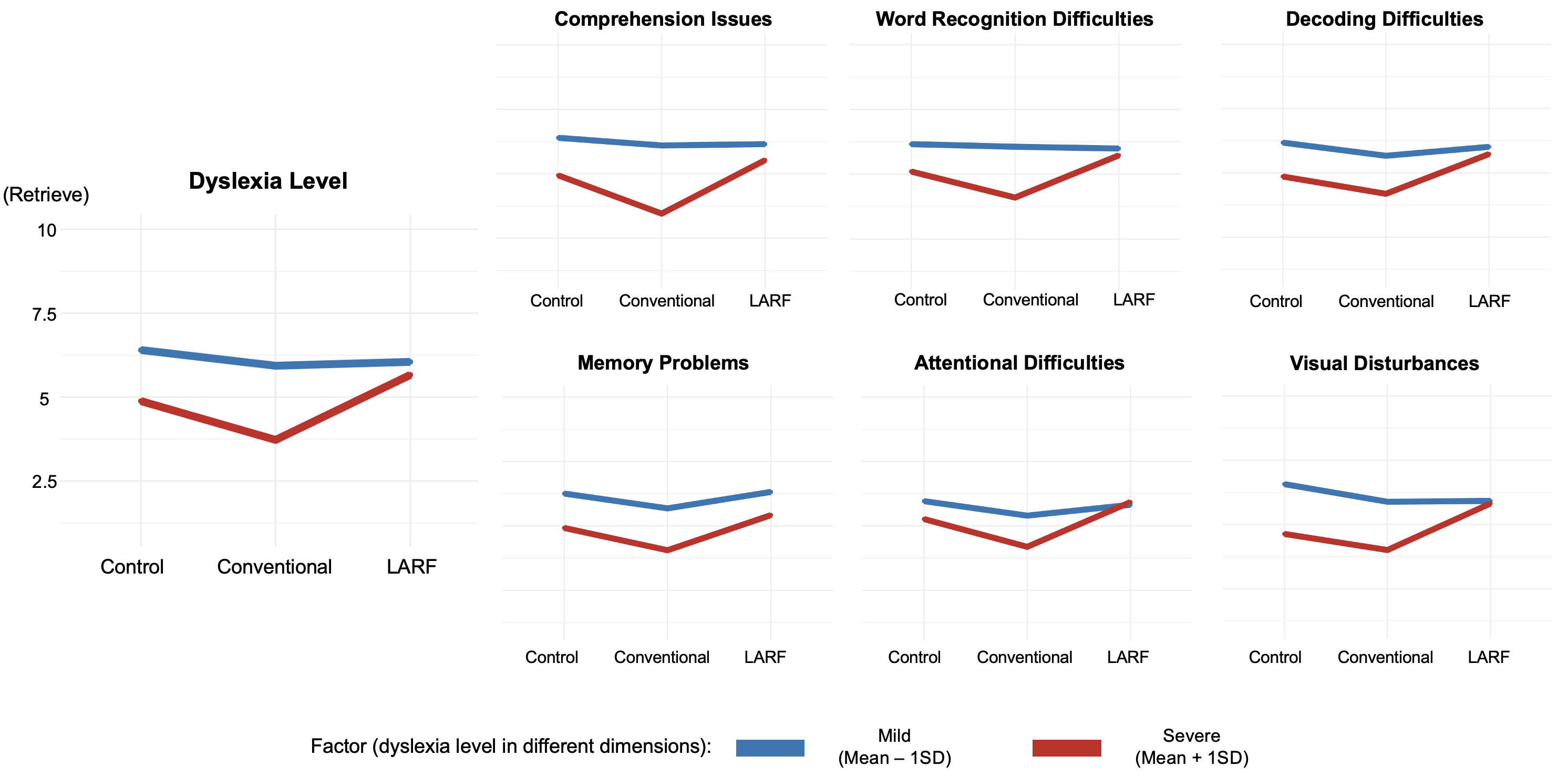}
  \Description{This figure presents six small line charts arranged in a 2×3 grid, each chart corresponding to a specific dyslexia-related dimension: “Comprehension Issues,” “Word Recognition Difficulties,” “Decoding Difficulties,” “Memory Problems,” “Attentional Difficulties,” and “Visual Disturbances.” Along the x-axis of each chart are three groups (Control, Conventional, LARF), and the y-axis is labeled “Dyslexia Level (Retrieve),” ranging roughly from 4 to 10. Two lines appear in each plot: one for mild dyslexia (blue) and one for severe dyslexia (red), each connecting the mean values (±1 SD) across the three groups. Typically, the red line (severe dyslexia) is higher than the blue line (mild dyslexia), reflecting greater reported difficulty among those with severe dyslexia. The line shapes vary by dimension; for example, in “Comprehension Issues,” the mild line forms a V shape from Control to Conventional to LARF, while the severe line has a slightly inverted pattern. Overall, these plots illustrate how participants with mild versus severe dyslexia perceive different levels of reading-related challenges under the three experimental conditions.}

  \caption{Post hoc evaluation for retrieving performance. The y-axis represents the scores for retrieve, with a maximum score of 10. While in the group with mild symptoms, LARF did not exhibit improvement, it significantly enhanced users' retrieval abilities in the group facing more severe reading challenges, whereas conventional tools had almost entirely negative impacts.}
  \label{fig:subfig retrieve}
\end{figure*}

\begin{figure*}[!htbp]
  \centering
  \includegraphics[width=0.8\textwidth]{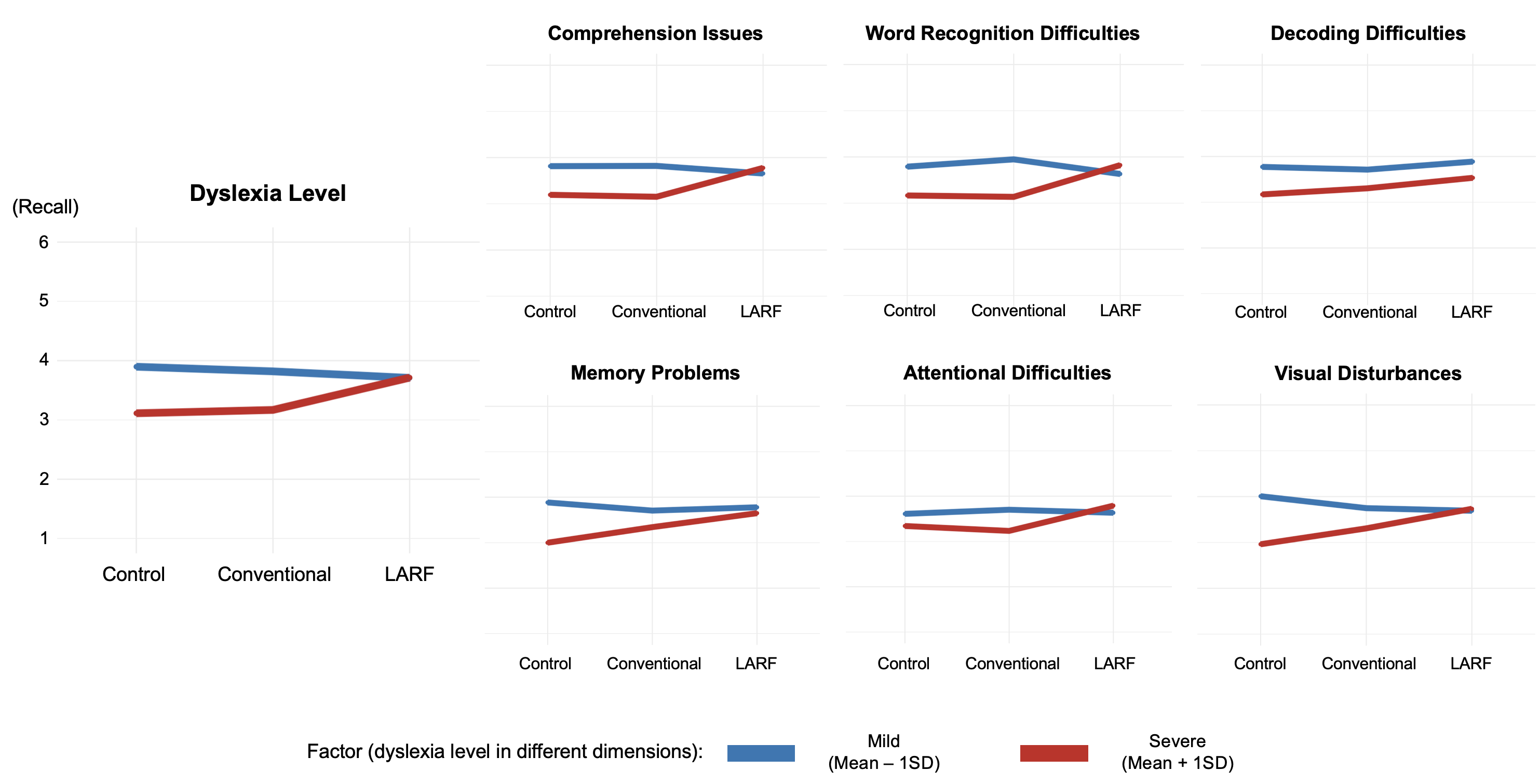}
  \Description{This figure contains six line charts arranged in a 2×3 grid, each labeled with a specific dyslexia-related dimension: “Comprehension Issues,” “Word Recognition Difficulties,” “Decoding Difficulties,” “Memory Problems,” “Attentional Difficulties,” and “Visual Disturbances.” Along the x-axis in each chart are three groups: Control, Conventional, and LARF. The y-axis, labeled “Dyslexia Level,” spans approximately from 1 to 6. Two lines are shown in each chart: one for participants with mild dyslexia (blue) and one for participants with severe dyslexia (red), each connecting mean values for the three groups, with “Mean ± 1 SD” indicated in the legend. Generally, the red (severe) line is higher than the blue (mild) line across all dimensions, reflecting greater difficulty levels for severe dyslexia. The differences among Control, Conventional, and LARF vary by dimension, illustrating how each group performs on these dyslexia-related factors.}

  \caption{Post hoc evaluation for recall performance. The y-axis represents the scores for recall, with a maximum score of 6. LARF similarly provided substantial assistance to the group with more severe symptoms, even surpassing the group with mild symptoms who also used LARF.}
  \label{fig:subfig recall}
\end{figure*}

\subsection{More Scenarios}
\begin{figure*}[!htbp]
  \centering
  \includegraphics[width=1\textwidth]{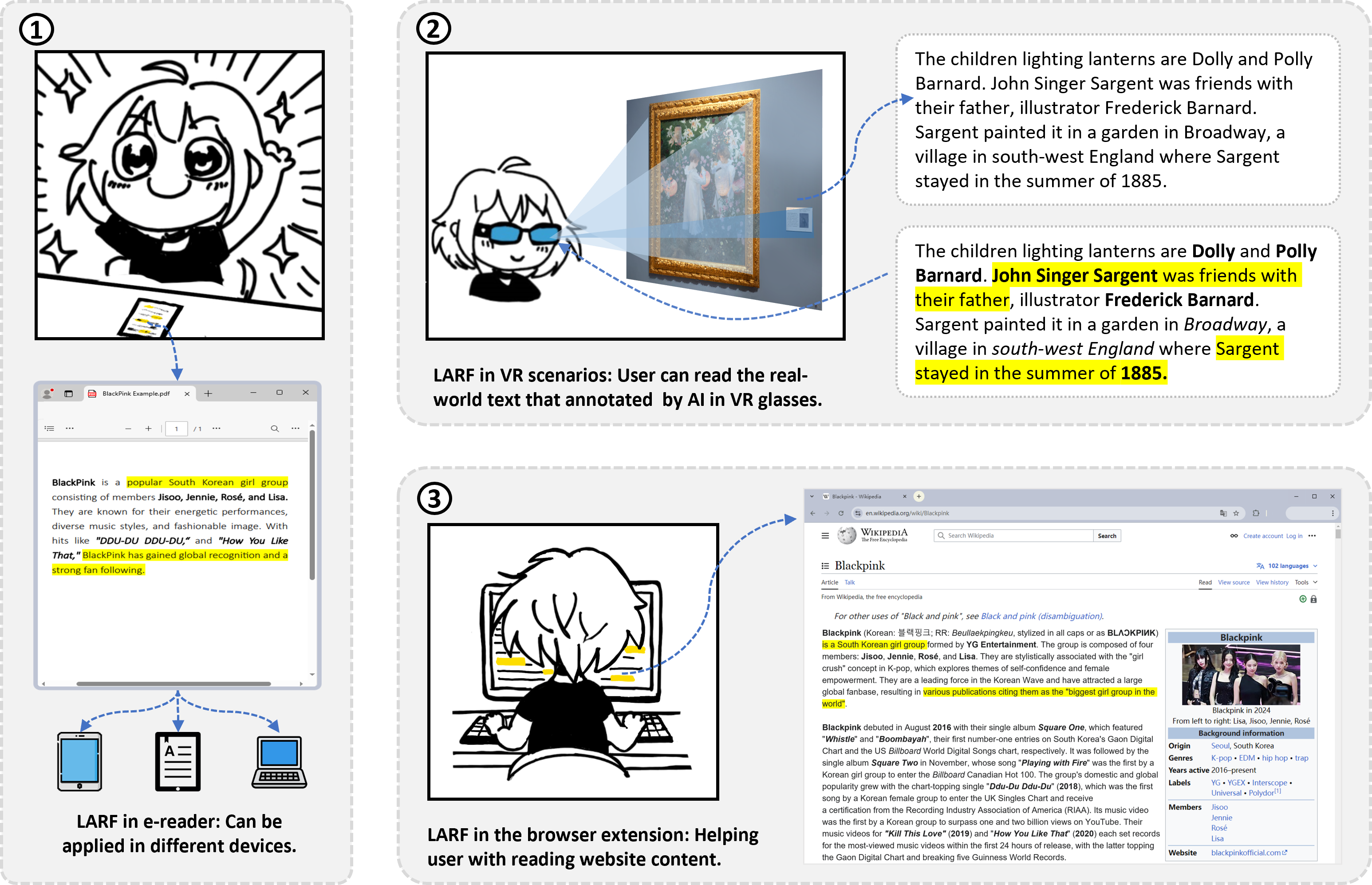}
  \Description{Real-world application scenarios that can apply LARF. In the first subplot, the user is using an e-reader which is integrated with LARF, this device can be a tablet, a smartphone or a laptop. In the second subplot, the user is wearing VR glasses, looking at the ``Carnation, Lily, Lily, Rose'' by John Singer Sargent. The VR headset with built-in LARF functionality helped annotate the description next to the painting, making it easier to read. In the Third subplot, LARF is integrated into a browser extension and helps the user reading the online content (the web page in the figure is the BlackPink item in Wikipedia}
  \caption{Real-world application scenarios that can apply LARF. In the first subplot, the user is using an e-reader which is integrated with LARF, this device can be a tablet, a smartphone or a laptop. In the second subplot, the user is wearing VR glasses, looking at the ``Carnation, Lily, Lily, Rose'' by John Singer Sargent \cite{sargentCarnation}. The VR headset with built-in LARF functionality helped annotate the description next to the painting, making it easier to read. In the Third subplot, LARF is integrated into a browser extension and helps the user reading the online content (the web page in the figure is the BlackPink item in Wikipedia\cite{blackpink2024}.)}
  \label{larf scenarios}
\end{figure*}

Besides LARF, there are lots of potential applications in different modalities and scenarios under the same idea: let AI decide and tell people what is worth attention. Some example is given in Figure \ref{larf scenarios}.
\end{document}